\documentclass[twocolumn,floatfix,superscriptaddress,showpacs,showkeys]{revtex4}
\makeatletter
\newcommand{\Rmnum}[1]{\expandafter\@slowromancap\romannumeral #1@}%大写罗马数字
\makeatother
\usepackage{amsmath}
\usepackage{booktabs}
\usepackage{amssymb}
\usepackage{multirow}
\usepackage{makecell}
\usepackage{graphicx}
\usepackage{textcomp}
\usepackage{subfigure}
\usepackage{phonetic}
\usepackage{extarrows}
\usepackage{color}
\usepackage{float}
\usepackage[colorlinks,citecolor=blue,linkcolor=blue]{hyperref}
\begin{document}
\title{Engineering of topological state transfer and topological beam splitter in an even-size Su-Schrieffer-Heeger chain}
\author{Lu Qi}
\affiliation{School of Physics, Harbin Institute of Technology, Harbin, Heilongjiang 150001, China}
\author{Guo-Li Wang}
\affiliation{School of Physics, Harbin Institute of Technology, Harbin, Heilongjiang 150001, China}
\author{Shutian Liu\footnote{E-mail: stliu@hit.edu.cn}}
\affiliation{School of Physics, Harbin Institute of Technology, Harbin, Heilongjiang 150001, China}
\author{Shou Zhang\footnote{E-mail: szhang@ybu.edu.cn}}
\affiliation{School of Physics, Harbin Institute of Technology, Harbin, Heilongjiang 150001, China}
\affiliation{Department of Physics, College of Science, Yanbian University, Yanji, Jilin 133002, China}
\author{Hong-Fu Wang\footnote{E-mail: hfwang@ybu.edu.cn}}
\affiliation{Department of Physics, College of Science, Yanbian University, Yanji, Jilin 133002, China}

\begin{abstract}
The usual Su-Schrieffer-Heeger model with an even number of lattice sites possesses two degenerate zero energy modes. The degeneracy of the zero energy modes leads to the mixing between the topological left and right edge states, which makes it difficult to implement the state transfer via topological edge channel. Here, enlightened by the Rice-Male topological pumping, we find that the staggered periodic next-nearest neighbor hoppings can also separate the initial mixed edge states, which ensures the state transfer between topological left and right edge states. Significantly, we construct an unique topological state transfer channel by introducing the staggered periodic on-site potentials and the periodic next-nearest neighbor hoppings added only on the odd sites simultaneously, and find that the state initially prepared at the last site can be transfered to the first two sites with the same probability distribution. This special topological state transfer channel is expected to realize a topological beam splitter, whose function is to make the initial photon at one position appear at two different positions with the same probability. Further, we demonstrate the feasibility of implementing the topological beam splitter based on the circuit quantum electrodynamic lattice. Our scheme opens up a new way for the realization of topological quantum information processing and provides a new path towards the engineering of new type of quantum optical device.   
     
\end{abstract}
\maketitle

\section{Introduction}\label{sec.1}
Topological insulator~\cite{hasan2010colloquium,qi2011topological,chiu2016classification,bansil2016colloquium}, as a new kind of novel state of the matters, has attracted increasing interest and attention since its identification. The significant difference between the traditional insulator and the topological insulator is the topologically inequivalent energy band structures in the momentum space~\cite{bansil2016colloquium,matsuura2010momentum}. The topological nonequivalence leads that the topological insulator holds the conducting edge states and the insulating bulk states at the same time~\cite{hasan2010colloquium,qi2011topological,chiu2016classification,bansil2016colloquium}. The special conducting edge states are protected by the energy gap of the topological system, leading the edge states to be immune to the local disorder and perturbation~\cite{hasan2010colloquium,qi2011topological,wray2011topological,xu2006stability,malki2017tunable,xiao2017observation,brouwer2011topological}. These novel properties support that the topological insulator has numerous latent applications in quantum information processing~\cite{alicea2011non,duclos2010fast} and the quantum computing~\cite{sarma2015majorana,stern2013topological} since these quantum issues both need to resist the deleterious effects of the local decoherence. The robust quantum state transfer assisted via the topological edge channel can be achieved based on the dipolar arrays~\cite{dlaska2017robust}. Similarly, the global fault-tolerant topological quantum computation can be constructed based on the non-local Majorana fermions~\cite{aasen2016milestones,sarma2015majorana,tewari2007quantum}.

As one of the simplest topological insulator model, the Su-Schrieffer-Heeger (SSH) model~\cite{su1979solitons,takayama1980continuum} has also attracted more and more attention in recent years since it possesses the structural simplicity and the abundant physical insights concurrently~\cite{fradkin1983phase,jackiw1976solitons,heeger1988solitons}. The structural simplicity ensures that the SSH model can be mapped by dint of all kinds of different systems, such as the cold atoms trapped into the optical lattice~\cite{meier2016observation,ganeshan2013topological,li2014topological,zhang2018topological}, the waveguide arrays~\cite{cheng2015topologically,longhi2013zak,ke2019topological}, the graphene nanoribbons~\cite{groning2018engineering,ribeiro2015transport,da2015impurity,cao2017topological}, the superconducting resonators and qubits~\cite{engelhardt2017topologically,paraoanu2014recent,qi2018simulation}, the optomechanical array composed by multiple single optomechanical system~\cite{esmann2018topological,xing2018controllable,xu2018generalized,qi2020controllable}, etc. Based on these platforms, various topological contexts in SSH model have been explored, including the edge state and the topological phase transition~\cite{grusdt2013topological,li2014topological,yao2018edge,di2016two}, quantum walk~\cite{kitagawa2010exploring,asboth2013bulk,ezawa2019electric}, the non-Hermitian effect~\cite{yao2018edge,lieu2018topological,zhu2014pt,kunst2018biorthogonal}, topological charge pumping~\cite{wang2013topological,lohse2016thouless,mei2018topology,hu2019dispersion}, the observation and detection of the topological features~\cite{meier2016observation,xie2019topological,rakovszky2017detecting,velasco2017realizing}, etc. Especially, in the context of the SSH model, the state transfer between the topological left and right edge states based on a superconducting qubit chains has been reported~\cite{mei2018robust}. Note that the superconducting-qubit-based SSH chain in Ref.~\cite{mei2018robust} was conceived to have an odd number of the lattice sites since the state transfer between the left and right edge states needs to occupy the same type of sites. Another reason is that the two edge modes of the SSH model with an even number of lattice sites were shown to be degenerate in the topologically non-trivial regions, which leads the left and right edge states are mixed. Thus, the state transfer between the left and right edge states cannot be achieved easily based on the even-size SSH model.           

In this paper, to surmount the obstacle mentioned above, we investigate several different topological state transfer processes based on a periodically modulated SSH model with an even number of the lattice sites. We first recall the Rice-Male (RM) topological pumping, in which, by adding the staggered periodic on-site potentials into the system, the initial degenerate zero energy modes in even-size SSH completely split and the topological state transfer between the left and right edge states can be realized. Enlightened by RM topological pumping, we find that the initial degenerate zero energy modes can also be split via introducing the staggered periodic next-nearest neighbor (NNN) hoppings since the NNN hoppings have the identical interaction forms as on-site energy in momentum space. In this way, the initial mixed left and right edge states are separated, which can be acted as the topologically protected edge channel to implement the state transfer between different sites. Specially, we find that, via introducing the staggered periodic on-site potentials and the periodic NNN hoppings added only on the odd sites simultaneously, a special topological state transfer channel can be opened. By dint of this topological edge channel, we find that the photon initially prepared at the rightmost site can be transferred to the first two sites with the same probability distribution, which implies the potential possibility of realizing a photon split-flow device, such as topological beam splitter. We demonstrate the feasibility of implementing the topological beam splitter assisted by the topological channel in detail, and find that the topological beam splitter is immune to the mild random disorders added into the system. Meanwhile, we propose to construct the topological beam splitter based on circuit quantum electrodynamic (circuit-QED) lattice, and show that it can be realized under the current experimental conditions. Our scheme greatly enriches the potential application of topological matter in quantum information processing and provides a new path towards the engineering of new type of quantum optical device. 
 
The paper is organized as follows: In Sec.~\ref{sec.2}, we mainly demonstrate the state transfer between the left and right edge states based on the even-size SSH model by introducing the staggered periodic NNN hoppings. In Sec.~\ref{sec.3}, we construct a special topological edge channel by the on-site potentials and the NNN hoppings added only on the odd sites. Meanwhile, we discuss the potential possibility of implementing topological beam splitter and show that it is immune to the mild disorder. Further, we demonstrate that the topological beam splitter can be experimentally realized based on circuit-QED lattice. Finally, a conclusion is given in Sec.~\ref{sec.4}.

\section{Topological state transfer induced by different parameter regimes}\label{sec.2}
\begin{figure}
	\centering
	\includegraphics[width=0.95\linewidth]{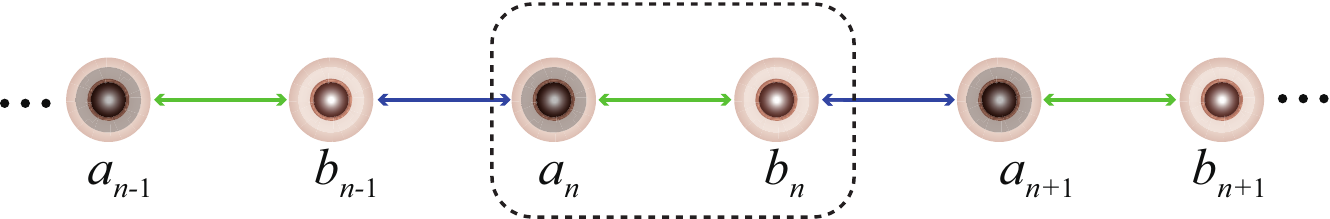}\\
	\caption{The diagrammatic sketch of the even-size SSH model. The SSH model is composed by $N$ unit cells (the black dashed rectangle), in which each unit cell contains an $a$-type site and a $b$-type site simultaneously. The size of the SSH model is $2N$.}\label{fig1}
\end{figure}
\begin{figure}
	\centering
	\subfigure{\includegraphics[width=0.75\linewidth]{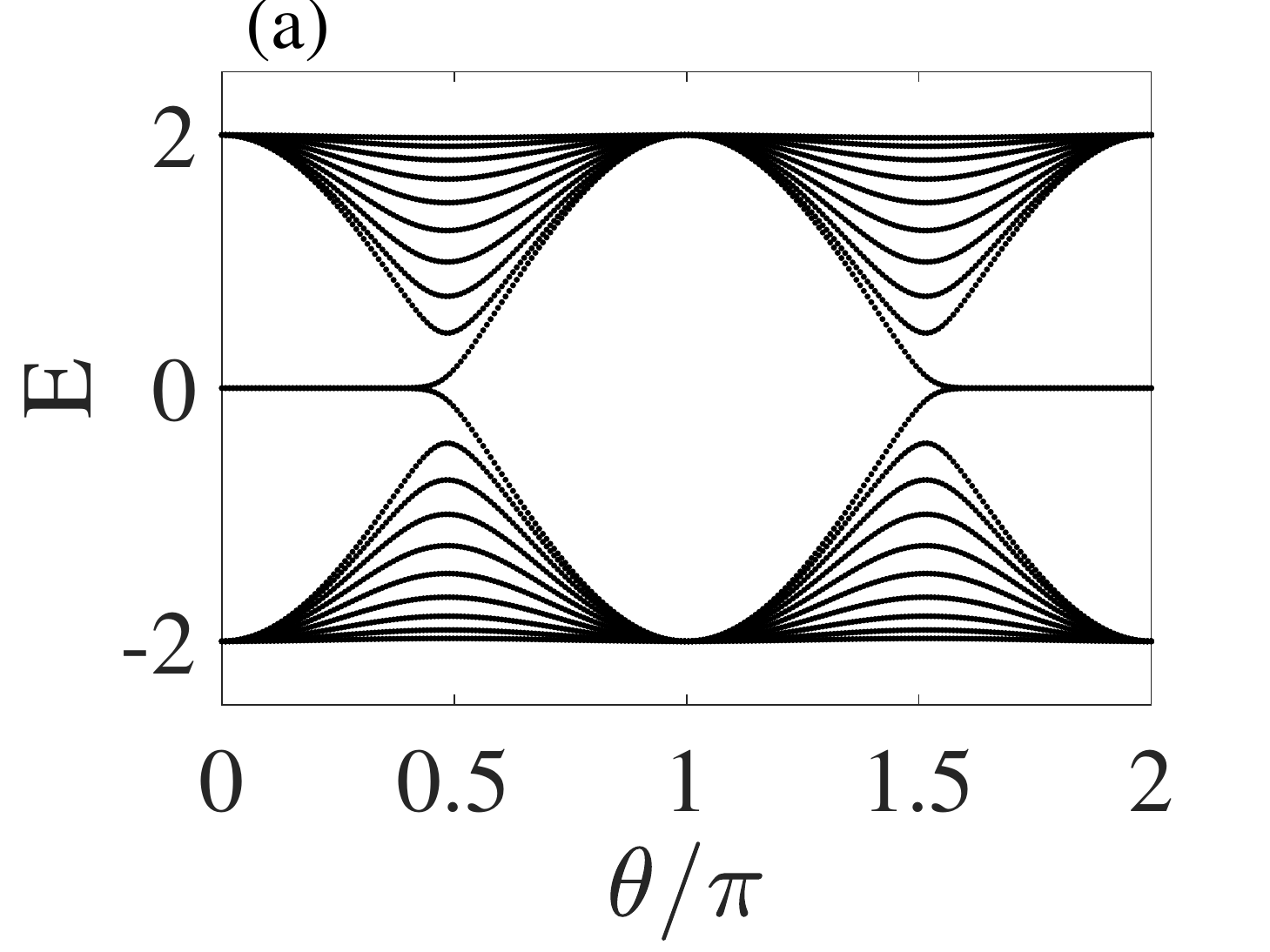}}
	
	\subfigure{\includegraphics[width=0.75\linewidth]{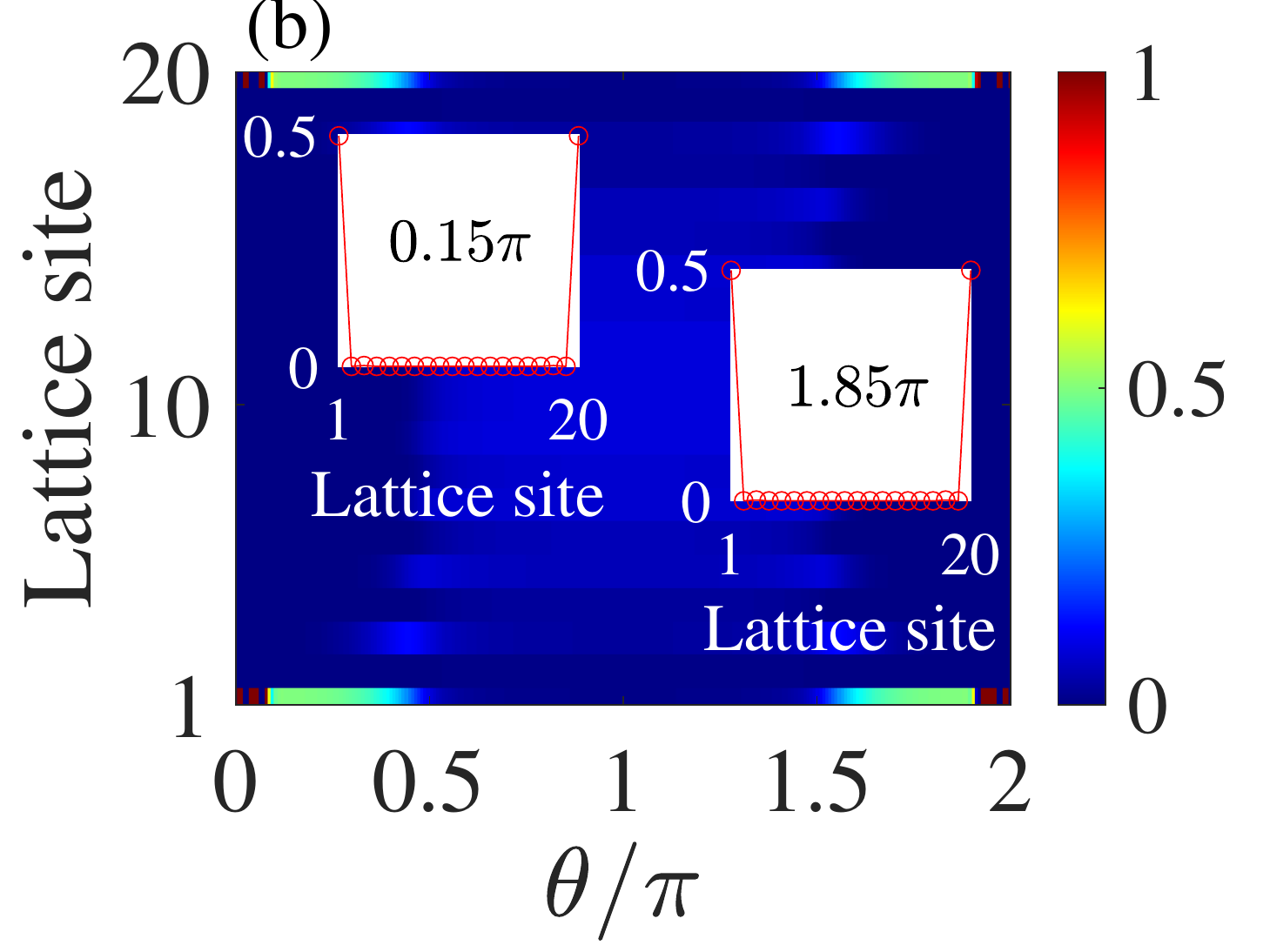}}
	\caption{The energy spectrum and the distribution of the zero energy mode. (a) The energy spectrum of the even-size SSH model. The energy gap has two degenerate zero energy modes. (b) The distribution of one zero energy mode [($N/2+1$)th zero energy mode] versus $\theta$ and lattice site. The insets show the same distributions of the selected zero energy state for different $\theta$ with $\theta=0.15\pi$ for left inset and $\theta=1.85\pi$ for right inset. The size of the lattice is $2N$ with $N=10$.}\label{fig2}
\end{figure}

Consider an usual SSH model with $N$ unit cells, as shown in Fig.~\ref{fig1}. The Hamiltonian of the system can be written as
\begin{eqnarray}\label{e01}
H&=&\sum_{n}\left[t_{1} a_{n}^{\dag}b_{n}+t_{2}a_{n+1}^{\dag}b_{n}\right]+\mathrm{H.c.},
\end{eqnarray}  
where $t_{1}=1-\cos\theta$ ($t_{2}=1+\cos\theta$) is the periodic intra-cell (inter-cell) nearest-neighbor (NN) coupling of the SSH model with $\theta\in[0,~2\pi]$ being the periodic parameter (Note that in all of the text, we keep the hopping terms as $t_{1}=1-\cos\theta$ and $t_{2}=1+\cos\theta$). Obviously, the SSH model possesses two degenerate zero energy modes in the gap when parameter $\theta$ belongs to $[0,~0.5\pi]\cup[1.5\pi,~2\pi]$, as shown in Fig.~\ref{fig2}(a). To further evaluate the topological properties of the zero energy states, we plot the state distribution of one selected zero mode, as shown in Fig.~\ref{fig2}(b). The numerical results reveal that the zero energy mode occupies both the two ends with maximal distributions in the most regions of $\theta\in[0, ~0.5\pi]\cup[1.5\pi,~2\pi]$. The reason of the special distribution of the edge mode is the existence of the degenerate zero energy modes. The degeneracy of the zero energy modes leads the topological left and right edge states to be mixed as one eigenstate, which makes it difficult to realize the state transfer between the topological left and the right states since the zero energy mode always holds the identical distribution with $\theta$ varying from $0$ to $2\pi$, as revealed in Fig.~\ref{fig2}(b). 
\begin{figure}
	\centering
	\subfigure{\includegraphics[width=0.48\linewidth]{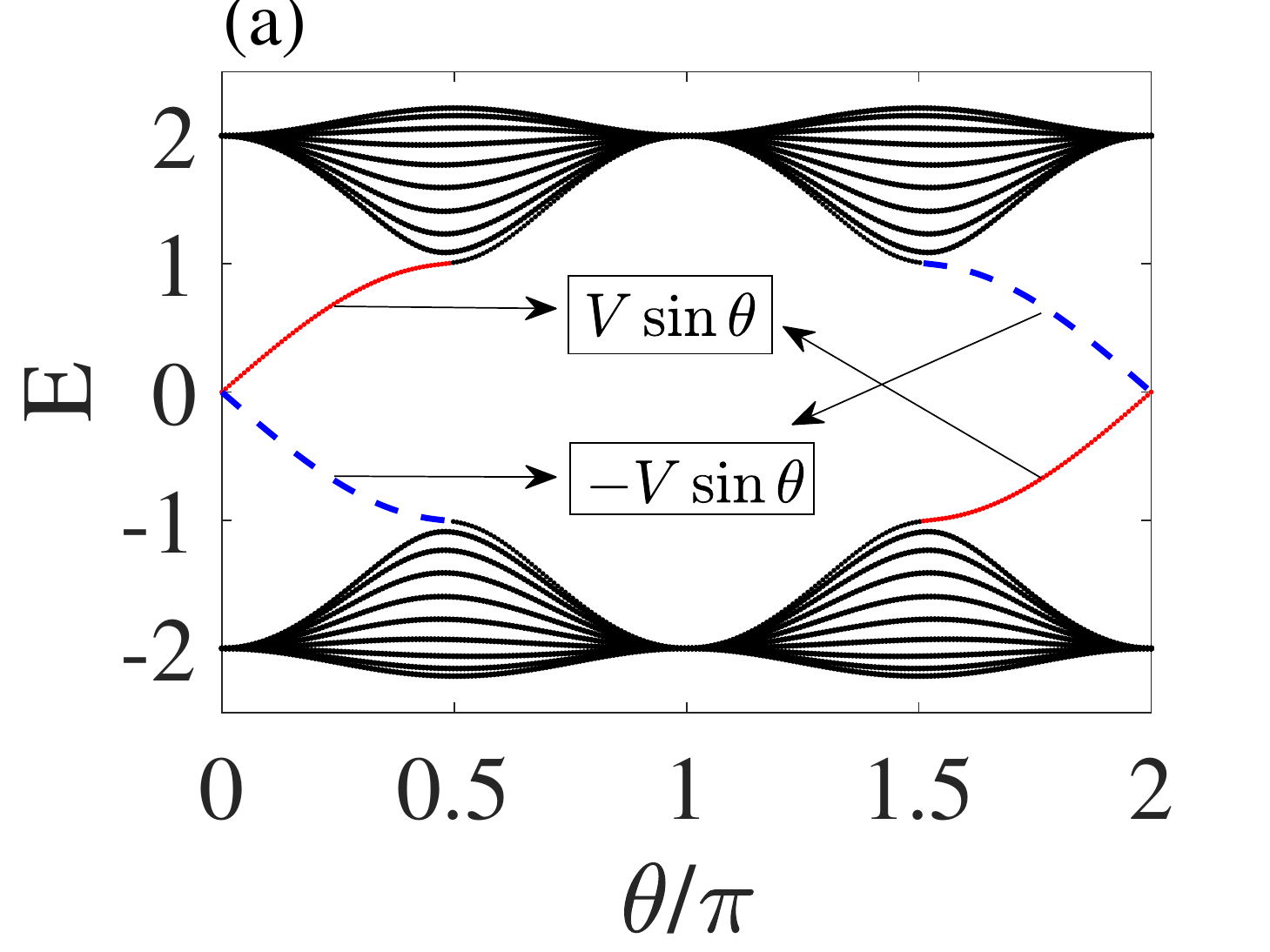}}
	\subfigure{\includegraphics[width=0.48\linewidth]{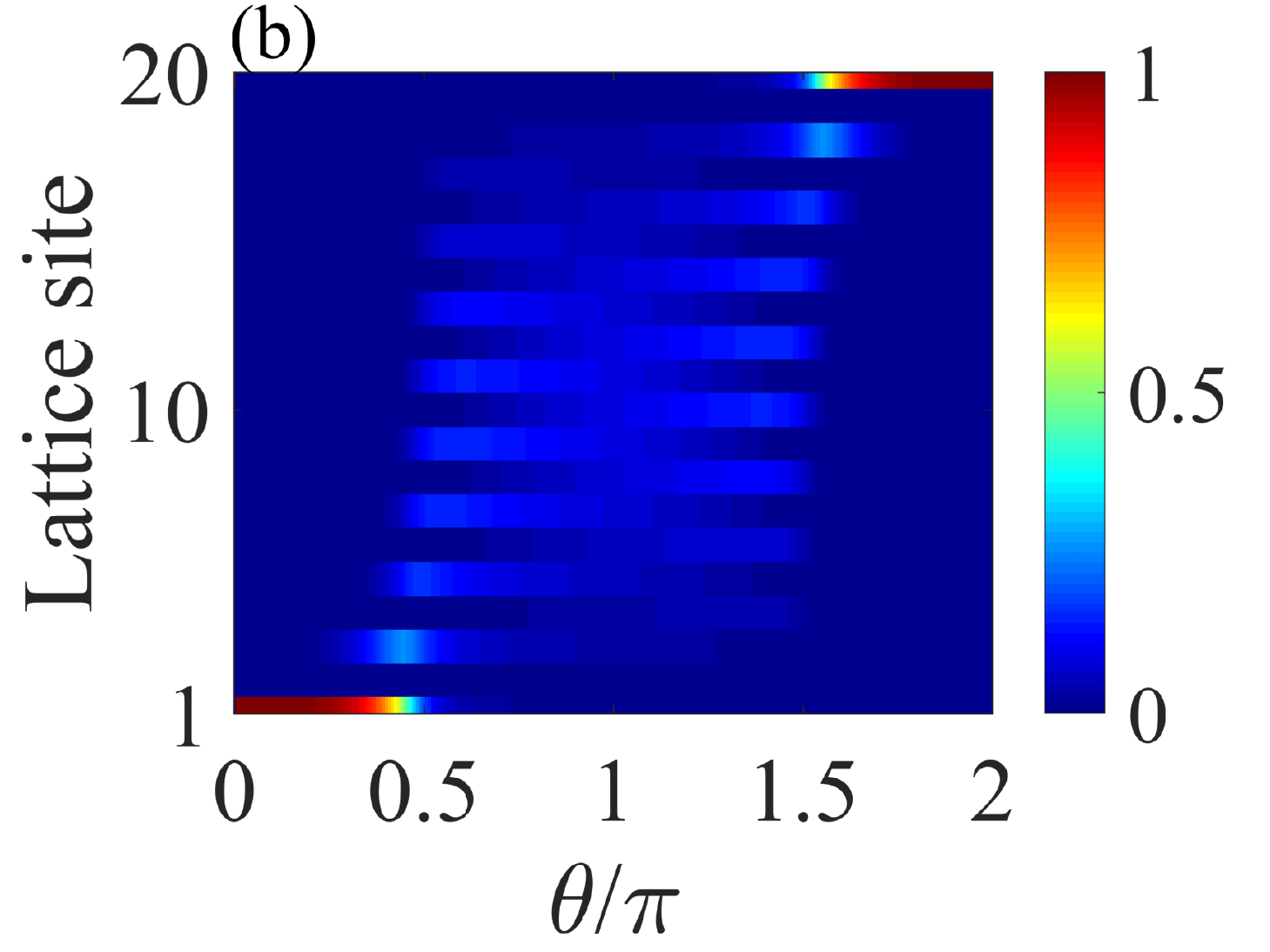}}
	
	\subfigure{\includegraphics[width=0.48\linewidth]{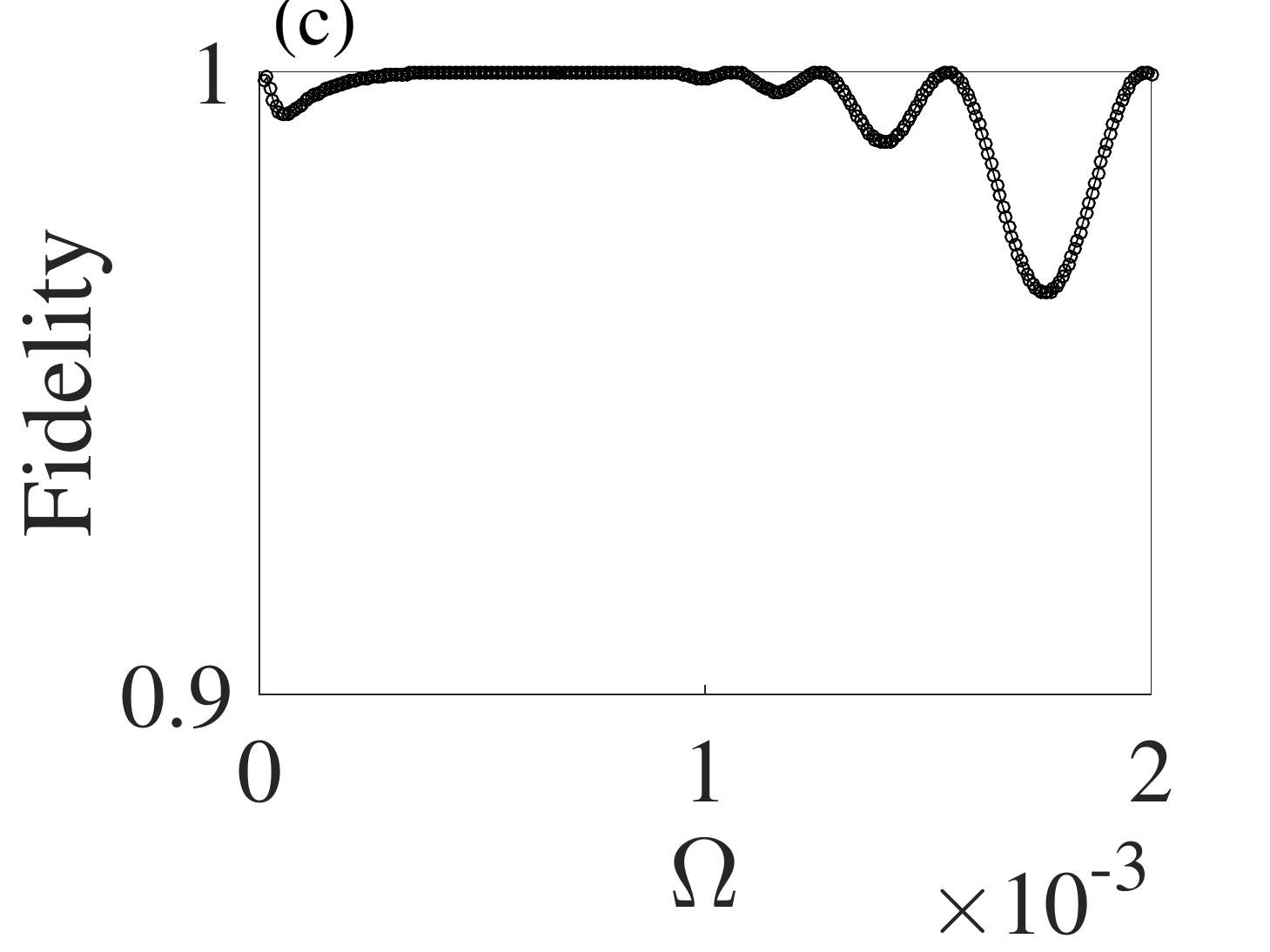}}
	\subfigure{\includegraphics[width=0.48\linewidth]{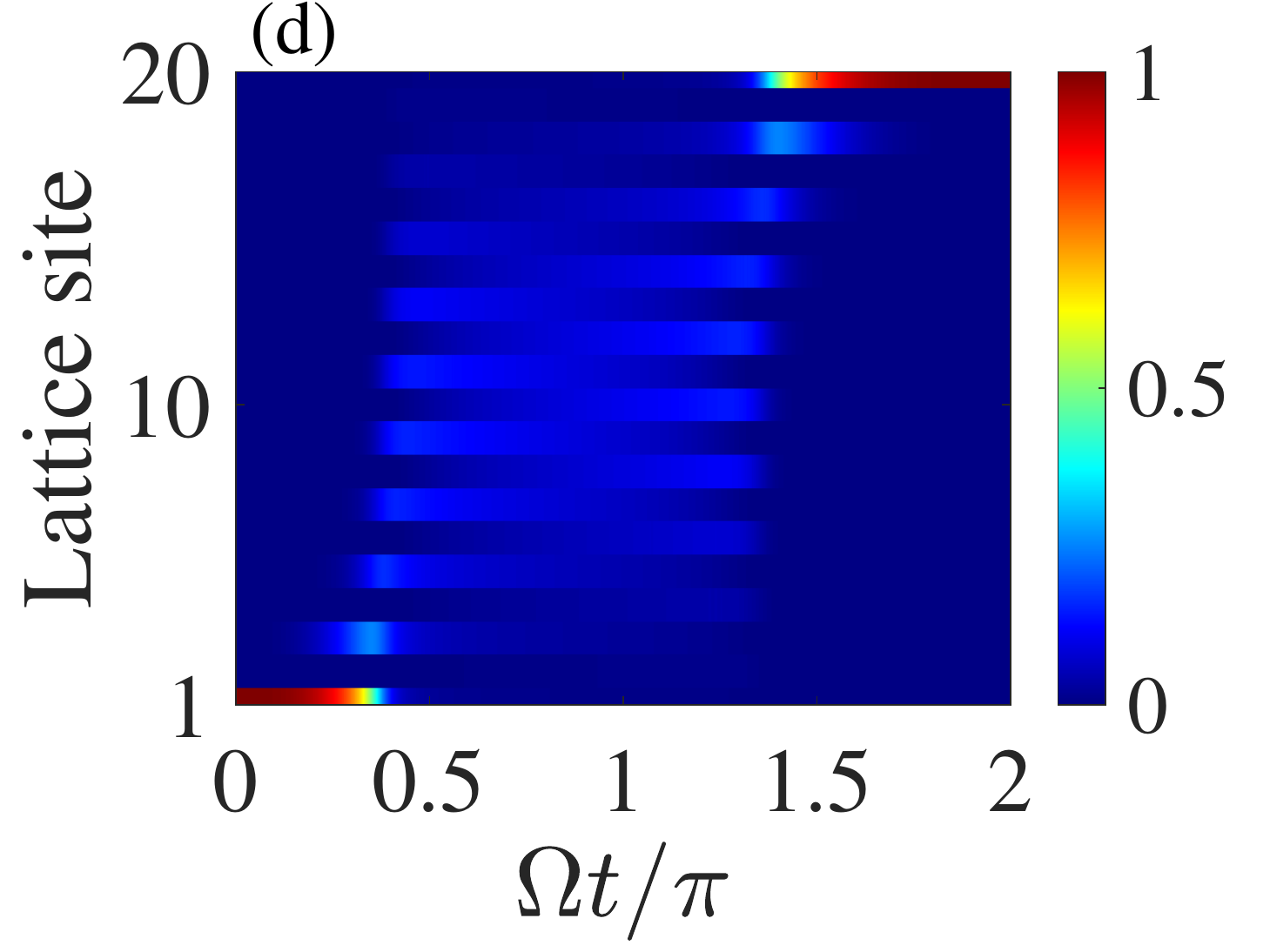}}
	\caption{(a) The energy spectrum of the SSH model when the periodic on-site potentials are added on $a$-type and $b$-type sites alternatively. The energy gap has two separated gap states around $E=0$ line. The red line represents that the gap state is localized at left edge while the blue dashed line represents that the gap state is localized at right edge. (b) The corresponding distribution of the upper gap state in (a). (c) The fidelity of the state transfer between $|1,~0,~0,...,0,~0\rangle$ and $|0,~0,~0,...,0,~1\rangle$ versus the varying rate of $\theta$. (d) The state transfer process corresponding to $\Omega=0.0005$. Other parameter takes $N=10$.}\label{fig3}
\end{figure} 
\begin{figure}
	\centering
	\subfigure{\includegraphics[width=0.75\linewidth]{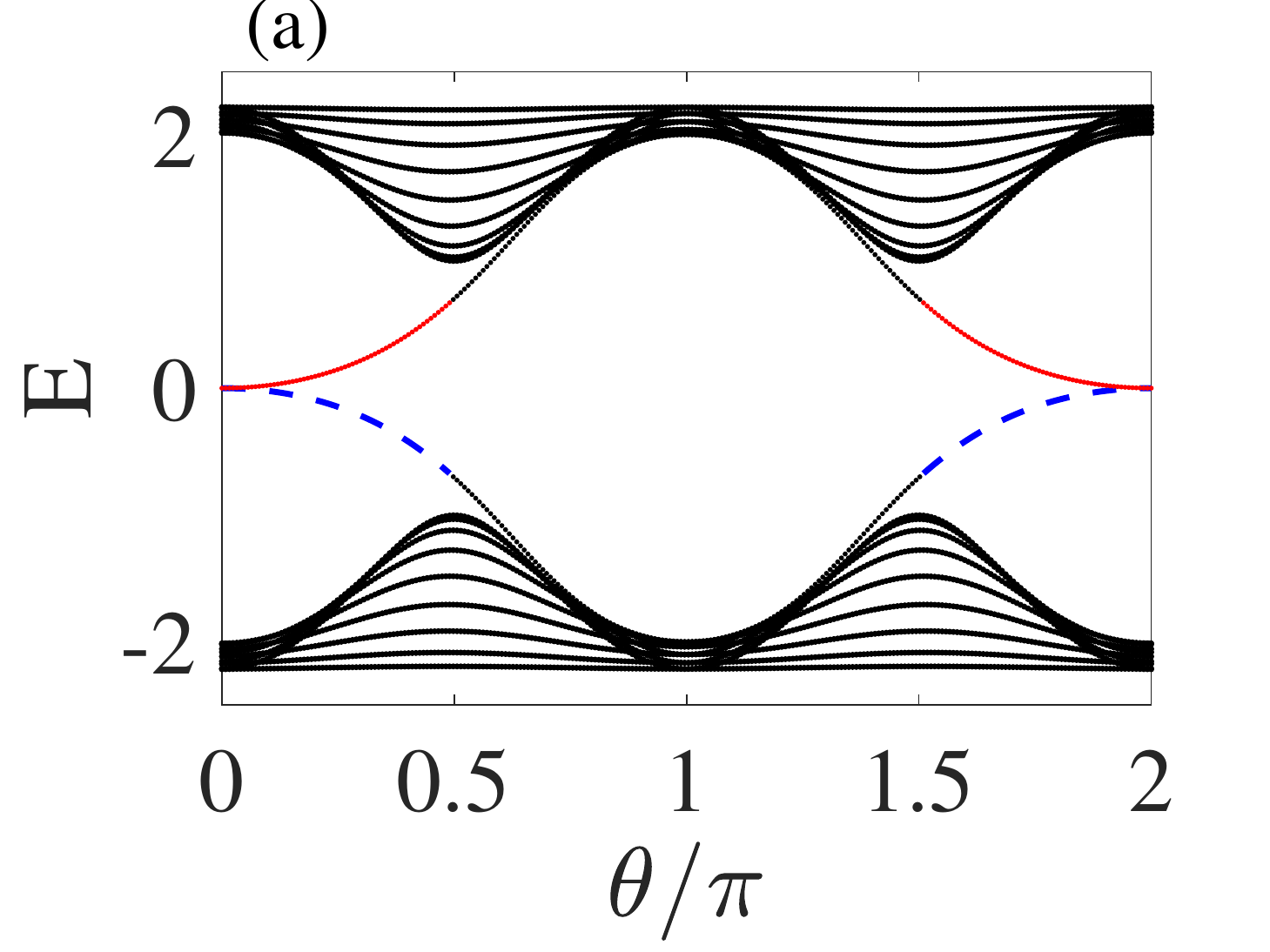}}
	
	\subfigure{\includegraphics[width=0.75\linewidth]{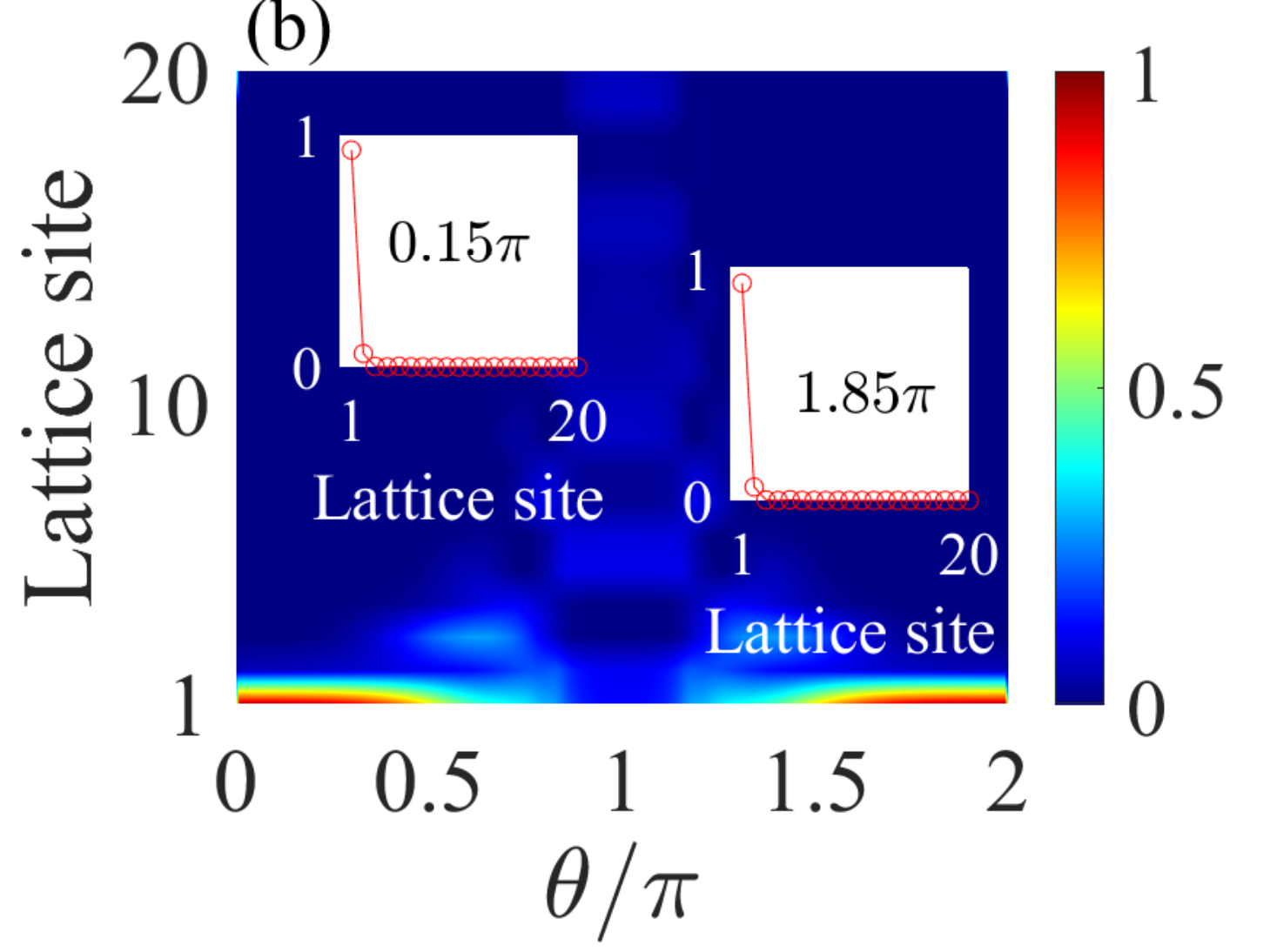}}
	\caption{The energy spectrum and the distribution of the gap state. (a) The energy spectrum of the system, where $T_{1}=-T_{2}=-0.5$. The energy gap has two separated gap states around $E=0$ line. The red line represents that the gap state is localized at left edge while the blue dashed line represents that the gap state is localized at right edge. (b) The distribution of the upper gap state in (a) versus $\theta$ and lattice site. The insets show the same distributions for different $\theta$ with $\theta=0.15\pi$ for left inste and $\theta=1.85\pi$ for right inset. The size of the lattice is $2N$ with $N=10$.}\label{fig4}
\end{figure}
\begin{figure}
	\centering
	\subfigure{\includegraphics[width=0.48\linewidth]{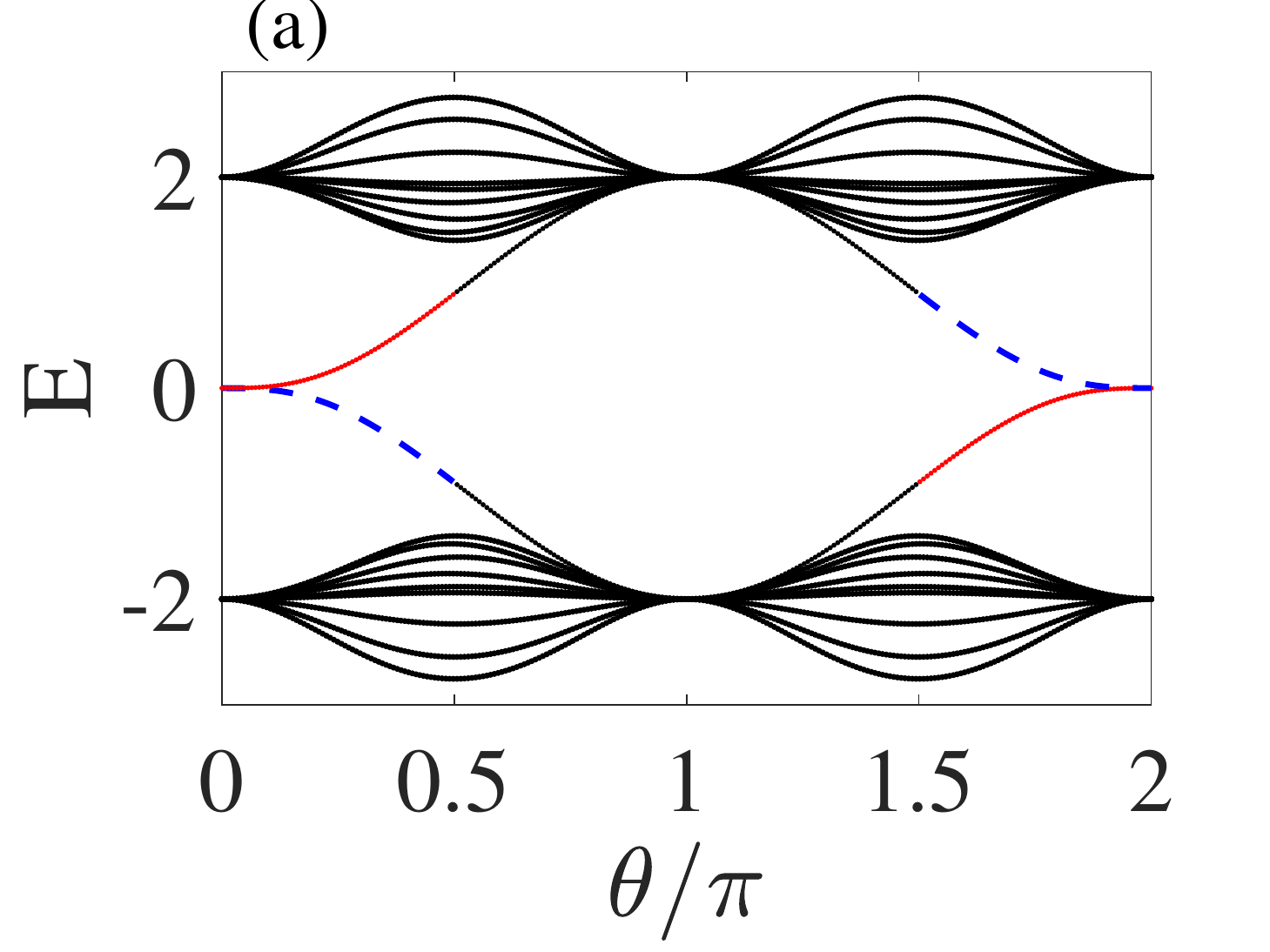}}
	\subfigure{\includegraphics[width=0.48\linewidth]{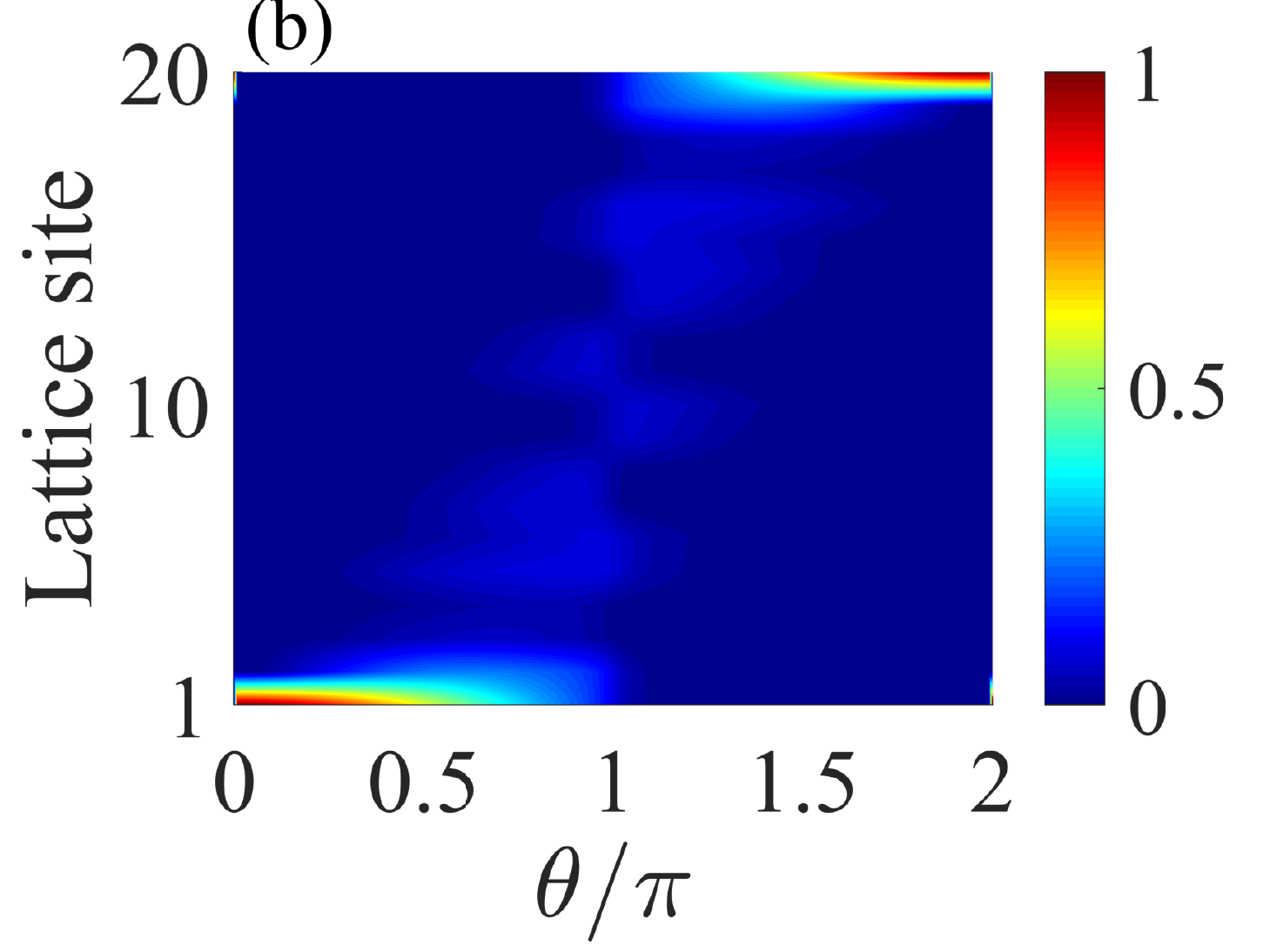}}
	
	\subfigure{\includegraphics[width=0.48\linewidth]{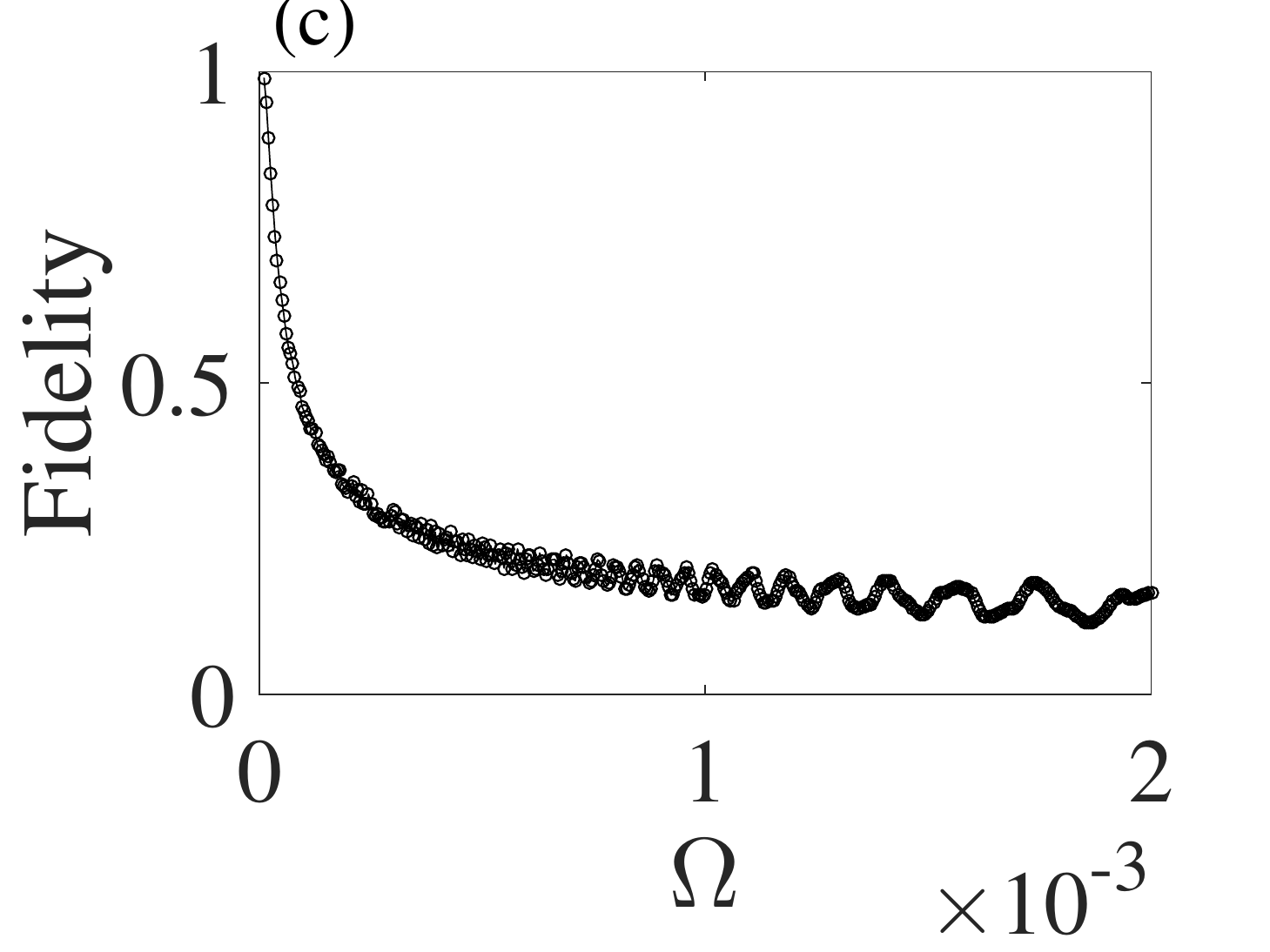}}
	\subfigure{\includegraphics[width=0.48\linewidth]{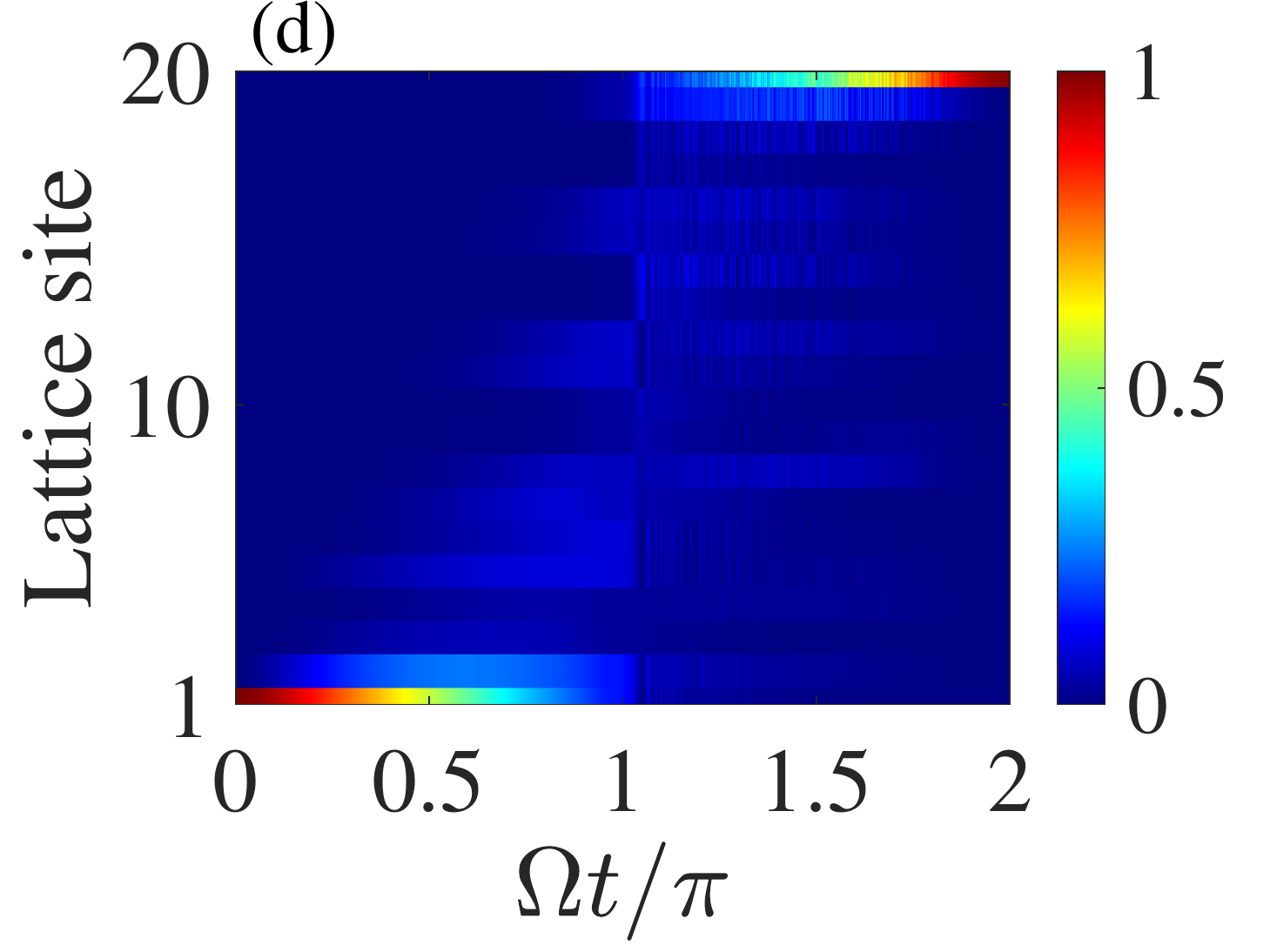}}
	\caption{(a) The energy spectrum of the SSH model when the periodic NNN hoppings $T_{1}=-\sin\theta$ and $T_{2}=\sin\theta$ are added into the system. The energy gap has two separated gap states around $E=0$ line. The red line represents that the gap state is localized at left edge while the blue dashed line represents that the gap state is localized at right edge. (b) The corresponding distribution of the upper gap state in (a). (c) The fidelity of the state transfer between $|1,~0,~0,...,0,~0\rangle$ and $|0,~0,~0,...,0,~1\rangle$ versus the varying rate of $\theta$. (d) The state transfer process corresponding to $\Omega=0.00001$. Other parameter takes $N=10$.}\label{fig5}
\end{figure}

\subsection{Topological state transfer induced by the staggered periodic on-site potentials}\label{sec.2A}
For the SSH model with the even number of the lattice sites, the degenerate zero energy modes make it impossible to realize the state transfer between the topological left and right edge states. We thus need to introduce different parameter regimes to break the degeneracy of the zero energy modes. An effective way is that we introduce the staggered periodic on-site potentials into the $a$-type and the $b$-type sites alternatively to split the degenerate zero energy modes~\cite{asboth2016ashort}, with
\begin{eqnarray}\label{e02}
H&=&\sum_{n}\left[V_{a} a_{n}^{\dag}a_{n}+V_{b} b_{n}^{\dag}b_{n}+t_{1} a_{n}^{\dag}b_{n}+t_{2}a_{n+1}^{\dag}b_{n}\right]+\mathrm{H.c.},\cr&&
\end{eqnarray}  
where $V_{a}=V\sin\theta$ and $V_{b}=-V\sin\theta$ are the periodic on-site potentials added on the $a$-type and the $b$-type sites, with $V$ being the potential strength. The breaking of the chiral symmetry induced by the staggered periodic on-site potentials leads to that the initial degenerate zero energy modes are split into two non-zero gap modes with the energies of $V\sin\theta$ and $-V\sin\theta$ ($-V\sin\theta$ and $V\sin\theta$) when $\theta\in[0,~0.5\pi]$ and $\theta\in[1.5\pi,~2\pi]$, as shown in Fig.~\ref{fig3}(a). The splitting of the degenerate zero energy modes ensures that the energies of the initial left and right edge states decouple from each other. More specifically, if we take the edge states ansatz $|\Psi\rangle_{E=V_{a},V_{b}}=\sum_{n}\lambda^{n}\left[\alpha a_{n}^{\dag}+\beta b_{n}^{\dag}\right] |G\rangle$, with $\lambda$ being the localized indexes, $\alpha$ and $\beta$ being the probability amplitude of the gap state,  the energy eigenvalue equation can be written as~\cite{mei2018robust}
\begin{eqnarray}\label{e03}
V_{a}\lambda^{n}\alpha a_{n}^{\dag}|G\rangle+t_{2}\lambda^{n+1}\alpha b_{n}^{\dag}|G\rangle+t_{1}\lambda^{n}\alpha b_{n}^{\dag}|G\rangle\cr\cr=V\sin\theta\lambda_{a}^{n}\alpha a_{n}^{\dag}|G\rangle\cr\cr
V_{b}\lambda^{n}\beta b_{n}^{\dag}|G\rangle+t_{2}\lambda^{n-1}\beta a_{n}^{\dag}|G\rangle+t_{1}\lambda^{n}\beta a_{n}^{\dag}|G\rangle\cr\cr=-V\sin\theta\lambda_{b}^{n}\beta b_{n}^{\dag}|G\rangle.
\end{eqnarray} 
Then, the topological left  edge state (with $E=V\sin\theta$) and the right edge state (with $E=-V\sin\theta$) can be obtained, with
\begin{eqnarray}\label{e04}
|\Psi\rangle_{L}&=&\sum_{n}\left[(-\frac{t_{1}}{t_{2}})^{n}\alpha a_{n}^{\dag}\right] |G\rangle,\cr\cr
|\Psi\rangle_{R}&=&\sum_{n}\left[(-\frac{t_{2}}{t_{1}})^{n}\beta b_{n}^{\dag}\right] |G\rangle.
\end{eqnarray} 
The analytical result indicates that the upper gap state in Fig.~\ref{fig3}(a) is localized near the leftmost $a$-type site when $\theta\in[0,~0.5\pi]$, and near the rightmost $b$-type site when $\theta\in[1.5\pi,~2\pi]$ (since the energies of the red gap state are $V\sin\theta$ and $-V\sin\theta$ respectively), implying that the localization of the upper gap state experiences the transfer from left edge to right edge assisted via bulk with $\theta$ varying from $0$ to $2\pi$. To further verify the above results, we plot the distribution of the upper gap state, as shown in Fig.~\ref{fig3}(b). The numerical results indicate that, if we vary the parameter $\theta$ from $0$ to $2\pi$ with time, it is natural to construct a channel of the state transfer between the topological left and right edge states. Meanwhile, we stress that the state transfer channel mentioned above is actually the RM pumping~\cite{asboth2016ashort}, namely, realizing the topological pumping by the modulated on-site potentials. 

To further verify the feasibility of the topological state transfer, we rewrite the periodic parameter $\theta$ in Eq.~(\ref{e02}) as $\theta_{t}=\Omega t$, with $\Omega$ being the varying rate and $t$ being the time. In the usual processes of the state transfer, the parameter needs to vary adiabatically to ensure the high enough probability of success. Dramatically, in the present system, we find that the tiny enough $\Omega$ does not always correspond to a maximal successful probability, as shown in Fig.~\ref{fig3}(c). However, we always can find a finite $\Omega$ to maximize the fidelity of the state transfer. 

When the initial state is prepared in the perfect topological left edge state $|\Psi\rangle_{L}=|1\rangle_{a_{1}}\otimes|0\rangle_{b_{1}}\otimes|0\rangle_{a_{2}}\otimes...\otimes|0\rangle_{a_{N}}\otimes|0\rangle_{b_{N}}=|1,~0,~0,...,0,~0\rangle$, we use the time-dependent Hamiltonian to evolve the initial state with $i\frac{d}{dt}|\Psi\rangle_{L}=H(\theta_{t})|\Psi\rangle_{L}$. The corresponding process of the state transfer between the topological left and right edge states is shown in Fig.~\ref{fig3}(d), when $\Omega=0.0005$. The numerical results reveal that the state transfer between $|1,~0,~0,...,0,~0\rangle$ and $|0,~0,~0,...,0,~1\rangle$ can be realized with a high fidelity. Note that our scheme actually realizes the topological state transfer between the leftmost $a$-type and the rightmost $b$-type sites, which is different from the previous schemes based on the SSH model with an odd number of the lattice sites~\cite{qi2020controllable,mei2018robust}. 
\begin{figure}
	\centering
	\subfigure{\includegraphics[width=0.7\linewidth]{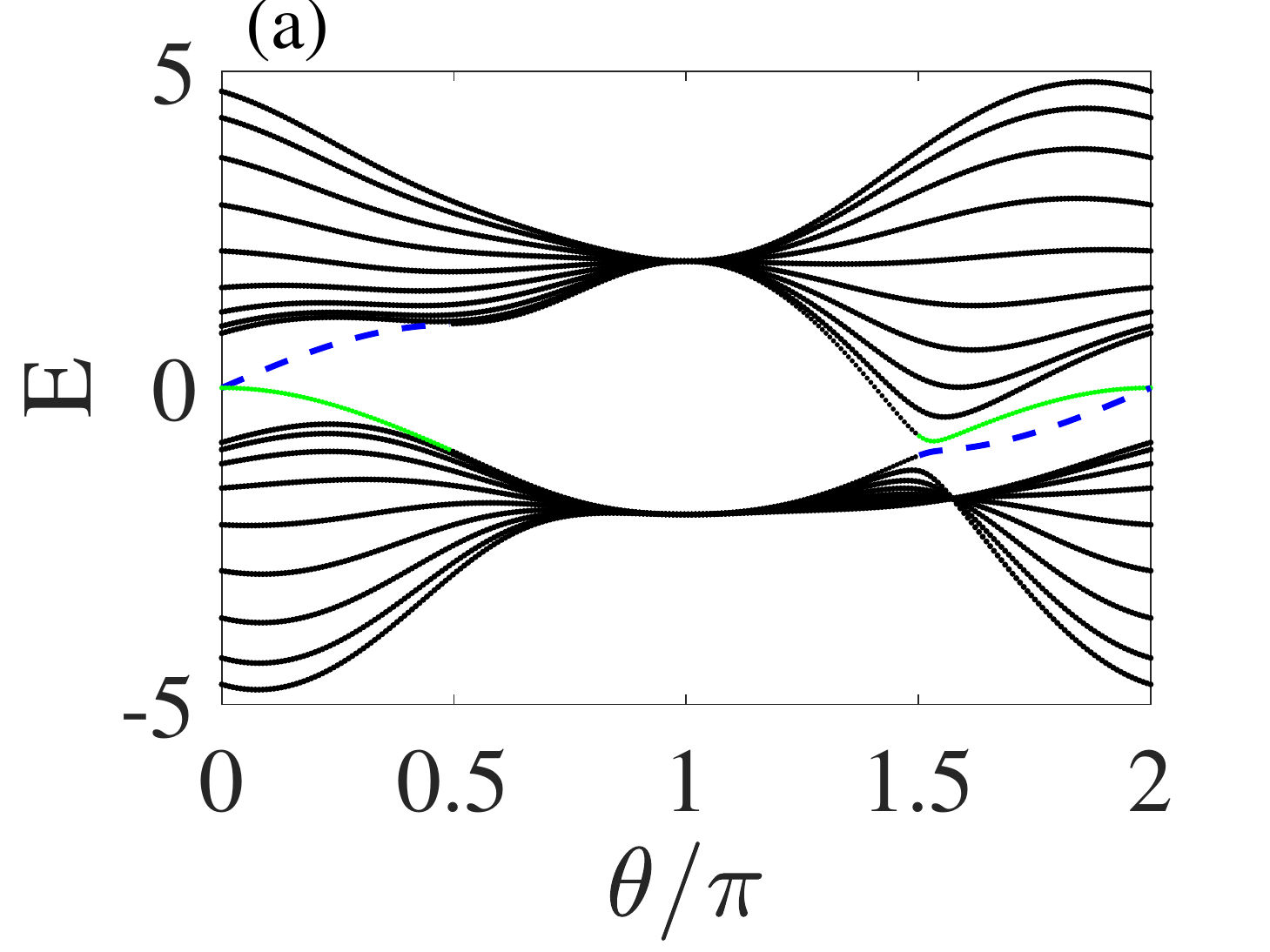}}
	
	\subfigure{\includegraphics[width=0.7\linewidth]{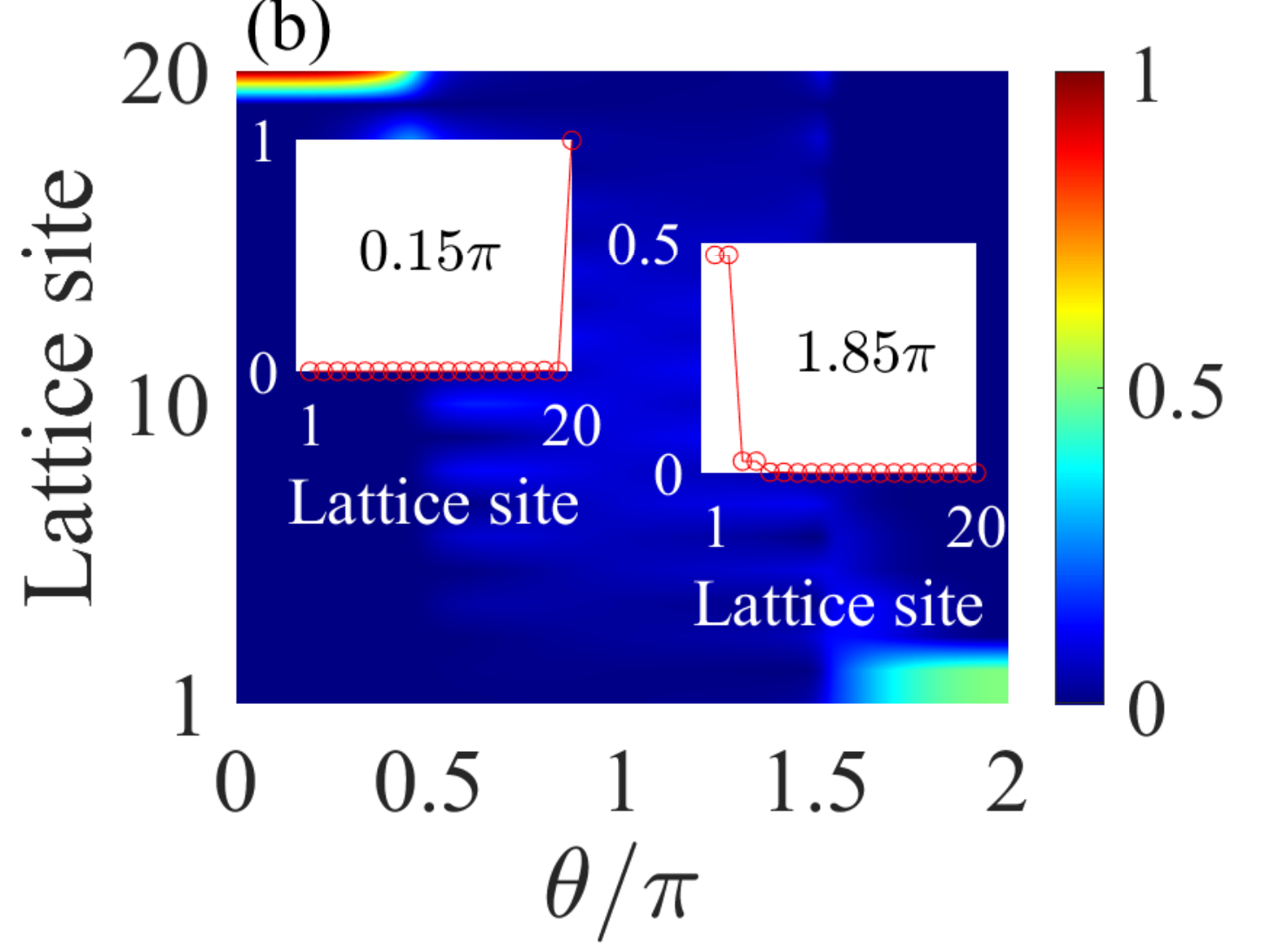}}
	
	\subfigure{\includegraphics[width=0.7\linewidth]{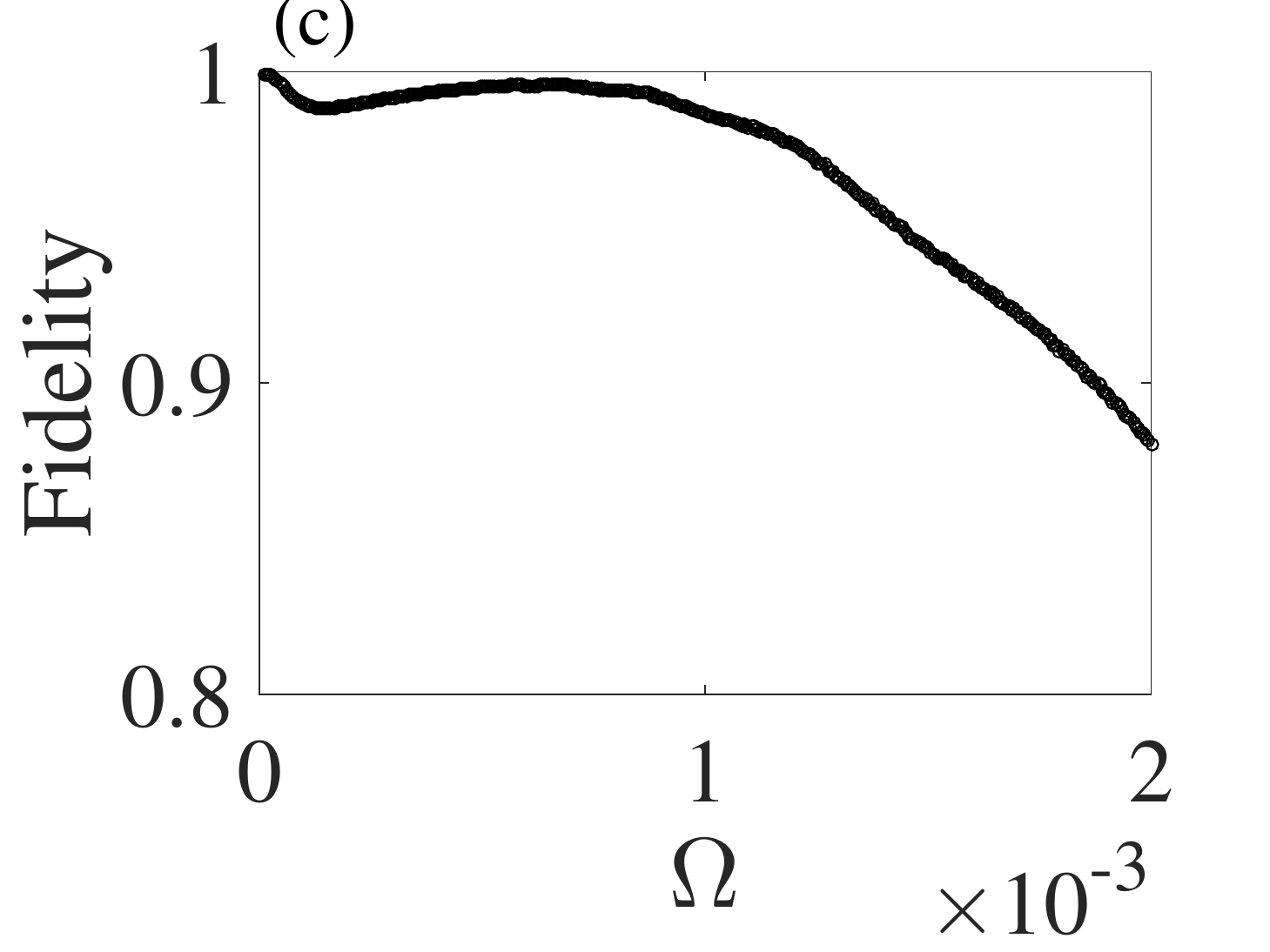}}
		
	\subfigure{\includegraphics[width=0.7\linewidth]{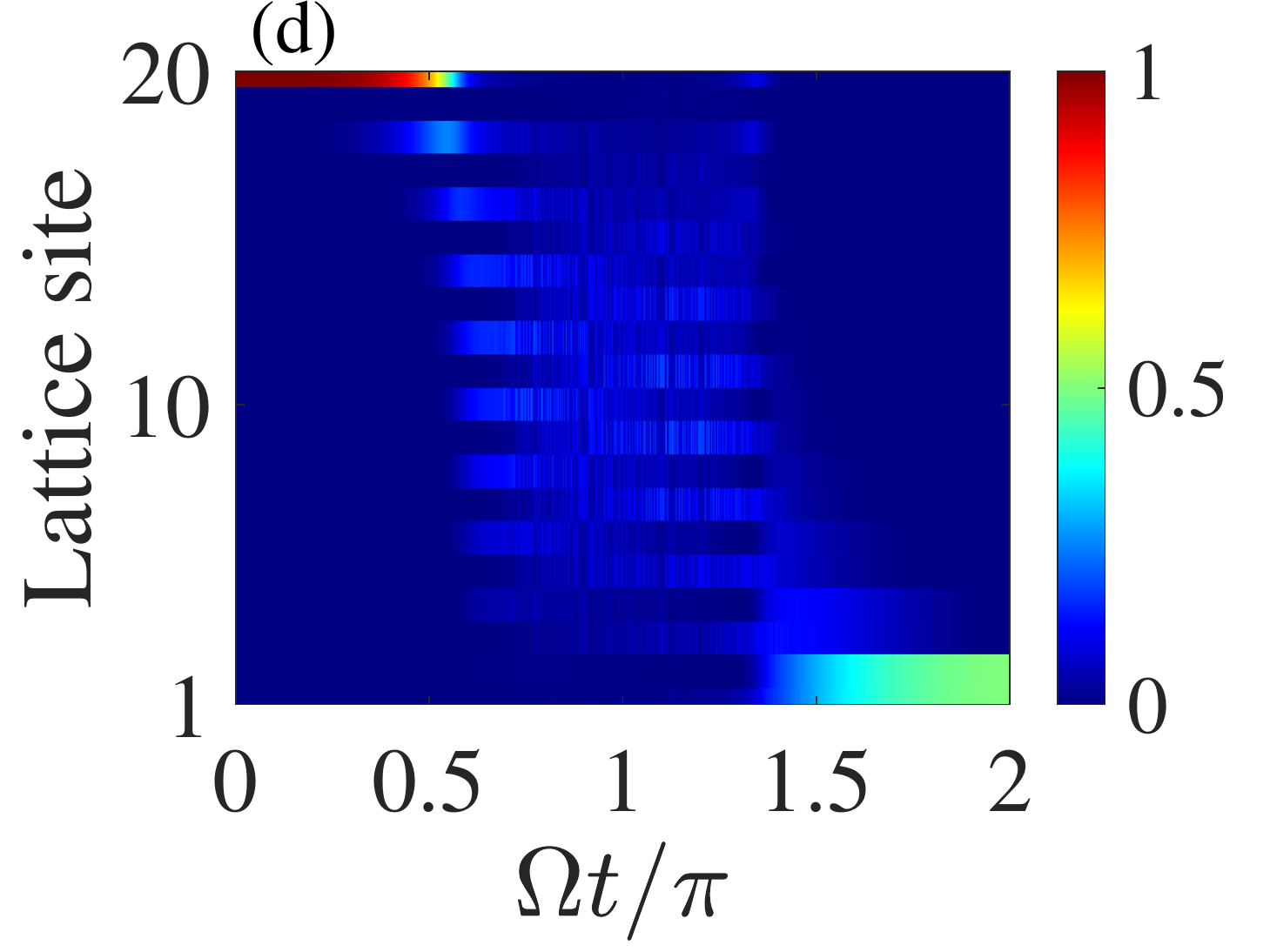}}
	\caption{(a) The energy spectrum of the SSH model when the periodic on-site potentials $V_{a}=-\sin\theta$ ($V_{b}=\sin\theta$) and the NNN hoppings $T_{1}=1+\cos\theta$ are added into system. The energy gap has two separated gap states. The blue dashed line represents that the gap state is localized at right edge while the green line represents that the gap state is mainly localized at first two sites with the same distributions. (b) The corresponding distribution of the upper gap state in (a). The insets show the distributions of the gap state for the different $\theta$ with $\theta=0.15\pi$ for left inset and $\theta=1.85\pi$ for right inset. (c) The fidelity of the state transfer between $|0,~0,~0,...,0,~1\rangle$ and $|0.5,~0.5,~0,...,0,~0\rangle$ versus the varying rate of $\theta$. (d) The state transfer process corresponding to $\Omega=0.00001$. Other parameter takes $N=10$.}\label{fig6}
\end{figure}

\subsection{Topological state transfer induced by the periodic next-nearest neighbor hopping}\label{sec.2B}
The essence of separating the degenerate zero energy modes is that the energies of the leftmost $a$-type site and the rightmost $b$-type site need to be shifted towards the opposite direction in the energy gap. Besides the RM pumping mentioned above, the energy shift of the two end sites can be achieved via introducing the NNN hoppings, with 
\begin{eqnarray}\label{e05}
H&=&\sum_{n}\left[t_{1} a_{n}^{\dag}b_{n}+t_{2}a_{n+1}^{\dag}b_{n}+T_{1}a_{n+1}^{\dag}a_{n}+T_{2}b_{n+1}^{\dag}b_{n}\right]\cr\cr
&&+\mathrm{H.c.},
\end{eqnarray}  
where $T_{1}$ and $T_{2}$ are the NNN hoppings strengths added on the $a$-type and the $b$-type sites. In this way, the energies of the two end sites can be shifted via the interactions of $T_{1}(a_{2}^{\dag}a_{1}+a_{1}^{\dag}a_{2})$ and $T_{2}(b_{N}^{\dag}b_{N-1}+b_{N-1}^{\dag}b_{N})$. We plot the energy spectrum of the present SSH model with the fixed values of the NNN hoppings, $T_{1}=-T_{2}=-0.5$, as shown in Fig.~\ref{fig4}(a). Obviously, the initial degenerate zero energy modes split in the most regions of $\theta\in[0,~0.5\pi]\cup[1.5\pi,~2\pi]$. We stress that, at the same time, the gap state still holds the same kind of the eigenstate when $\theta\in[0,~0.5\pi]\cup[1.5\pi,~ 2\pi]$. More specifically, as shown in Fig.~\ref{fig4}(a), the upper gap state is localized near the leftmost site both when $\theta\in[0,~0.5\pi]$ and $\theta\in[1.5\pi,~ 2\pi]$ while the bottom gap state is both localized near the rightmost site. To further clarify it, we simulate the distribution of the upper gap state in Fig.~\ref{fig4}(a) versus the parameter $\theta$ and the lattice site numerically, as shown in Fig.~\ref{fig4}(b). The results show that the upper gap state is indeed localized near the leftmost $a$-type site when $\theta\in[0,~0.5\pi]\cup[1.5\pi,~ 2\pi]$ except for the points of $\theta=0.5\pi$ and $\theta=1.5\pi$, at which the gap state integrates into the bulk.

The above results reveal that the fixed NNN hoppings cannot make the gap state own different distributions when $\theta\in[0,~0.5\pi]$ and $\theta\in[1.5\pi,~2\pi]$, which determines that the state transfer between the left and right edge states cannot be achieved via varying the periodic parameter adiabatically. Actually, the NNN hoppings play the analogous role as the on-site energy added on the $a$-type and the $b$-type sites, since the NNN hoppings are equivalent to the on-site energy in the momentum space with $T_{1}a_{k}^{\dag}a_{k}$ and $T_{2}b_{k}^{\dag}b_{k}$. Thus, utilizing the conclusions obtained in section~\ref{sec.2A} directly, we also assume that the NNN hoppings added on the $a$-type and the $b$-type sites varies with the parameter $\theta$ alternatively, with $T_{1}=-T_{2}=-V\sin\theta$. 

To verify the rationality of the assumption, we plot the energy spectrum and the distribution of the upper gap state, as shown in Figs.~\ref{fig5}(a) and \ref{fig5}(b). The numerical results reveal that, as a matter of fact, the periodic NNN hoppings ensure that the gap state has the opposite distributions when $\theta\in[0,~0.5\pi]$ and $\theta\in[1.5\pi,~ 2\pi]$, which provides the theoretical foundation to implement the state transfer between the topological left and right edge states. To evaluate the availability of the state transfer channel induced by the modulated NNN hoppings, we prepare the initial state in $|1,~0,~0,...,0,~0\rangle$ and plot the fidelity between the target state $|0,~0,~0,...,0,~1\rangle$ and the evolved final state, as shown in Fig.~\ref{fig5}(c). We find that, different from the case in section~\ref{sec.2A}, the fidelity decays rapidly with the increase of the varying rate $\Omega$. It means that we should take a small enough $\Omega$ to ensure the high enough probability of success for the state transfer. For example, when we take $\Omega=0.00001$ and prepare the initial state in $|1,~0,~0,...,0,~0\rangle$, the evolved final state is shown in Fig.~\ref{fig5}(d). The numerical results show that the state transfer between the left edge state and the right edge state can be achieved, meaning that the modulated NNN hoppings can indeed induce the topological state transfer channel. 
\begin{figure}
	\centering
	\includegraphics[width=1.0\linewidth]{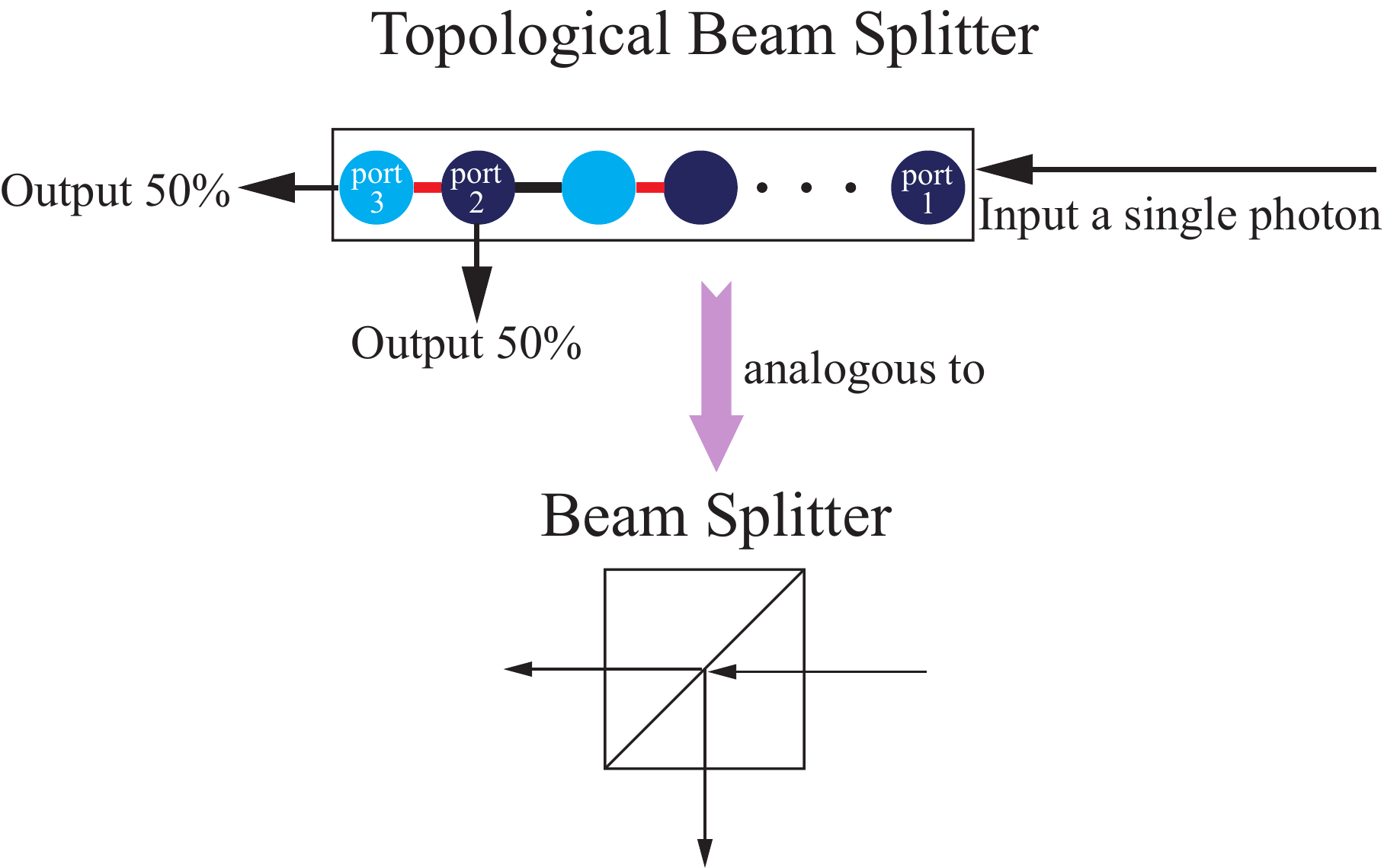}
	\caption{The sketch of the topological beam splitter. The topological beam splitter has three ports, in which the photons coming from the port $1$ finally gather into port $2$ and port $3$ with the same half of probability. From this perspective the present system is equivalent to a beam splitter [the bottom panel].}\label{fig7}
\end{figure}
\begin{figure}
	\centering
	\includegraphics[width=1.0\linewidth]{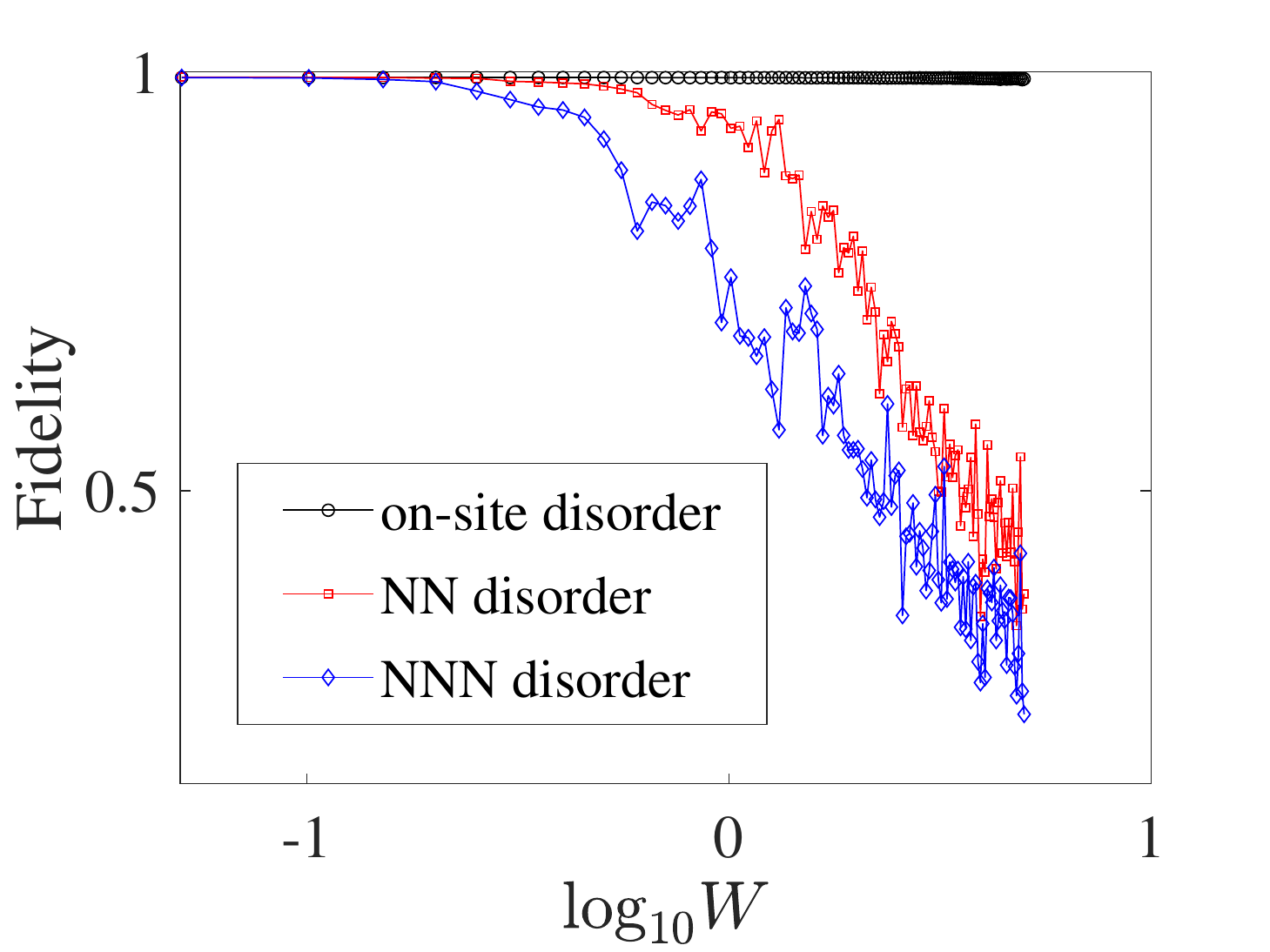}
	\caption{The fidelity of the topological beam splitter when the random disorder is added into the system. The black circle line represents that the random disorder is added into the modulated on-site energy $V_{a}$ and $V_{b}$. The red square line represents that the random disorder is added into the modulated NN hopping $t_{1}$ and $t_{2}$. The blue diamond line represents that the random disorder is added into the modulated NNN hoppings strength $T_{1}$. Note that the disorder is randomly sampled $100$ times and then taking the mean value.}\label{fig8}
\end{figure} 

\section{Topological beam splitter induced by the modulated on-site  potentials and the modulated partial NNN hoppings}\label{sec.3}
In section~\ref{sec.2}, we have shown the implementation of the state transfer between the left edge state and the right edge state by dint of the channels induced by the modulated on-site potentials and the modulated NNN hoppings, respectively. A question arises: can we realize other state transfer via designing the modulated on-site potentials and NNN hoppings? In the following, we show a special state transfer channel induced by the joint effects of the modulated on-site potentials and NNN hoppings. At the same time, we find that this special state transfer can be expected to achieve a topological beam splitter~\cite{Jonckheere2017Hanbury,Wang2017Topologically,Hammer2013Dynamics}, which provides the potential application towards topological quantum information processing.

\subsection{Topological beam splitter}\label{sec.3A}
Taking the modulated on-site potentials and NNN hoppings into account simultaneously, the Hamiltonian of the system is given by
\begin{eqnarray}\label{e06}
H&=&\sum_{n}\left[V_{a} a_{n}^{\dag}a_{n}+V_{b} b_{n}^{\dag}b_{n}+t_{1} a_{n}^{\dag}b_{n}+t_{2}a_{n+1}^{\dag}b_{n}\right.\cr\cr
&&\left.+T_{1}a_{n+1}^{\dag}a_{n}+T_{2}b_{n+1}^{\dag}b_{n}\right]+\mathrm{H.c.},
\end{eqnarray}  
where $V_{a}=-V_{b}=-\sin \theta$ is the modulated on-site energy, $t_{1}=1-\cos\theta$ and $t_{2}=1+\cos\theta$ are the modulated NN hopping terms, and $T_{1}=t_{2}=1+\cos\theta$ and $T_{2}=0$ represent the NNN hoppings added only on the odd sites. In this way, we find that the energy spectrum of the system becomes irregular but still has two gap sates, as shown in Fig.~\ref{fig6}(a). The corresponding distribution of the upper gap state in Fig.~\ref{fig6}(a) is shown in Fig.~\ref{fig6}(b). The numerical results reveal that the upper gap state exhibits a peculiar distribution, in which the gap state is mainly localized near the last site in $\theta\in[0, 0.5\pi]$ while it is mainly localized at the first two sites with the approximately same probability in $\theta\in[1.5\pi, 2\pi]$. It illuminates us that, by dint of the gap state induced by $V_{a}$, $V_{b}$, and $T_{1}$, we can achieve the special state transfer between the state $|0,~0,~0,...,0,~1\rangle$ and the state $|0.5,~0.5,~0,...,0,~0\rangle$. To further verify the above supposition, we rewrite the parameter $\theta$ as $\theta_{t}=\Omega t$ and use the time-dependent Hamiltonian to evolve the initial state prepared in $|0,~0,~0,...,0,~1\rangle$. The corresponding fidelity between the ideal state $|0.5,~0.5,~0,...,0,~0\rangle$ and the evolved final state versus the parameter $\Omega$ is shown in Fig.~\ref{fig6}(c). The numerical results reveal that we always can find an appropriate $\Omega$ to ensure the achievement of the state transfer between $|0,~0,~0,...,0,~1\rangle$ and $|0.5,~0.5,~0,...,0,~0\rangle$. For example, when $\Omega=0.00001$, we find that the photon initially in the last site can be transferred to the first two sites with the same distributions, as shown in Fig.~\ref{fig6}(d).    

The above results imply that, if we first inject a photon into the last site, after a certain time evolution, the photon will finally appear in the first two sites with the approximately same probability of $50\%$. In this way, from the perspective of regarding the last site and the first two sites as three ports, the probability of photon from port $1$ appearing in port $2$ and port $3$ is approximately the same, that is, $50\%$, as shown in Fig.~\ref{fig7}. Obviously, this phenomenon is analogous with the conventional beam splitter~\cite{Macneille1946,Schonbrun2006Yamashita,Chen2004Photonic,Donatella2000Beam} in quantum optics, showing that the present system can be expected to achieve an analogous beam splitter. Besides, we stress that the implementation of the above state transfer is assisted by the gap state, leading the process of the state transfer to be protected by the energy gap. Thus, the above state transfer between $|0,~0,~0,...,0,~1\rangle$ and $|0.5,~0.5,~0,...,0,~0\rangle$ can be used to map a topological beam splitter. 
The present topological beam splitter is naturally immune to the mild disorder and perturbation due to the protection of the gap. To clarify it, we consider a random disorder $W\delta$ is added into the system, with $W$ being the disorder strength and $\delta$ being the random number in the range of $[-0.5, 0.5]$. The success probability of the topological beam splitter, when the random disorder is added into the modulated on-site energy $V_{a}$ and $V_{b}$, the modulated NN hopping strengths $t_{1}$ and $t_{2}$, and the modulated NNN hoppings strength $T_{1}$, is depicted in Fig.~\ref{fig8}. The numerical results reveal that, for a certain range of the disorder strength with $\mathrm{log}_{10}W<-1$, the mild disorder has no effects on the topological beam splitter. The robustness of the topological beam splitter to the disorder and perturbation provides much more convenience for the experimental realization and the practical application of the topological beam splitter.  
 
\begin{figure}
	\centering
	\subfigure{\includegraphics[width=0.48\linewidth]{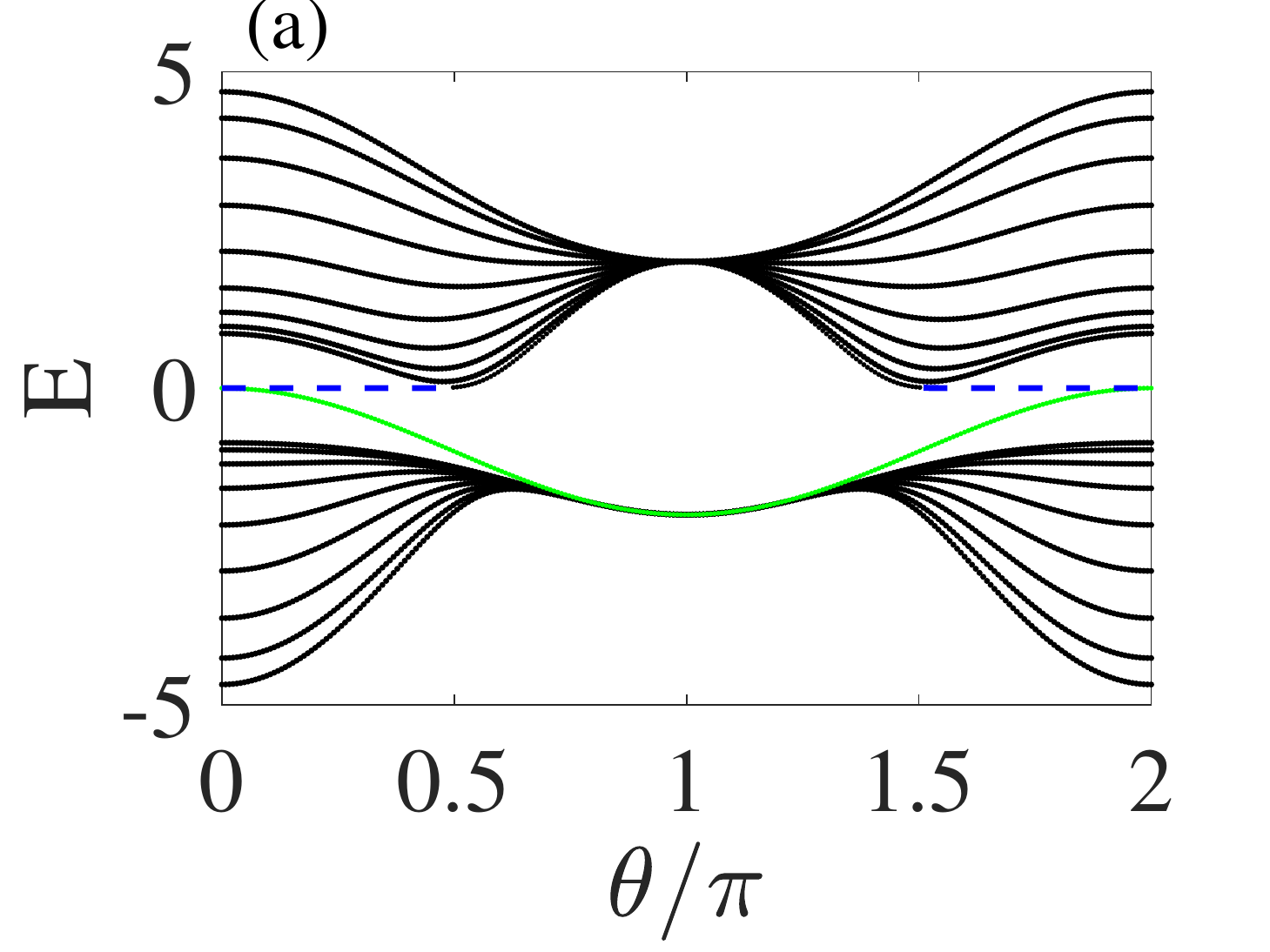}}
	\subfigure{\includegraphics[width=0.48\linewidth]{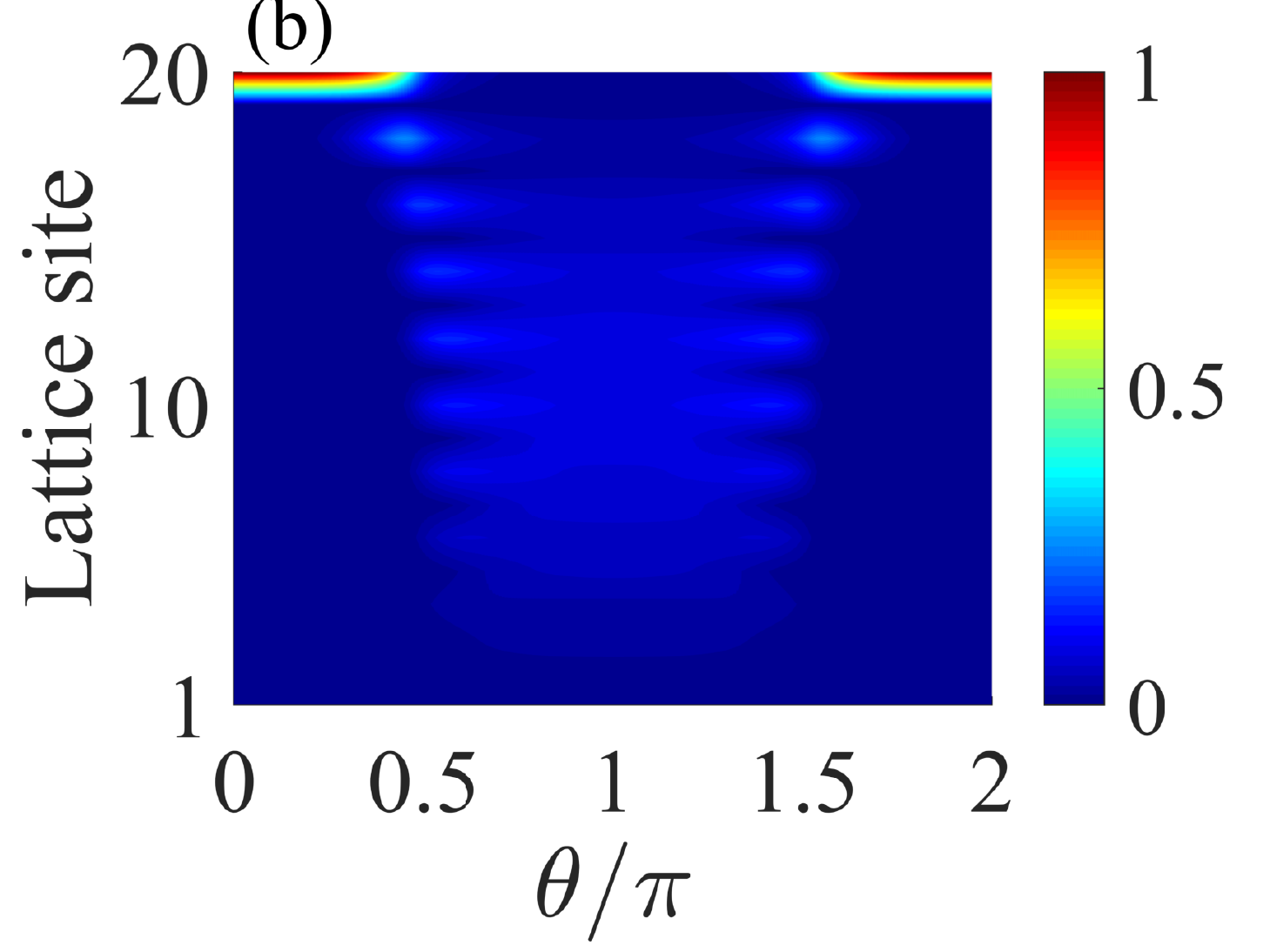}}
	
	\subfigure{\includegraphics[width=0.48\linewidth]{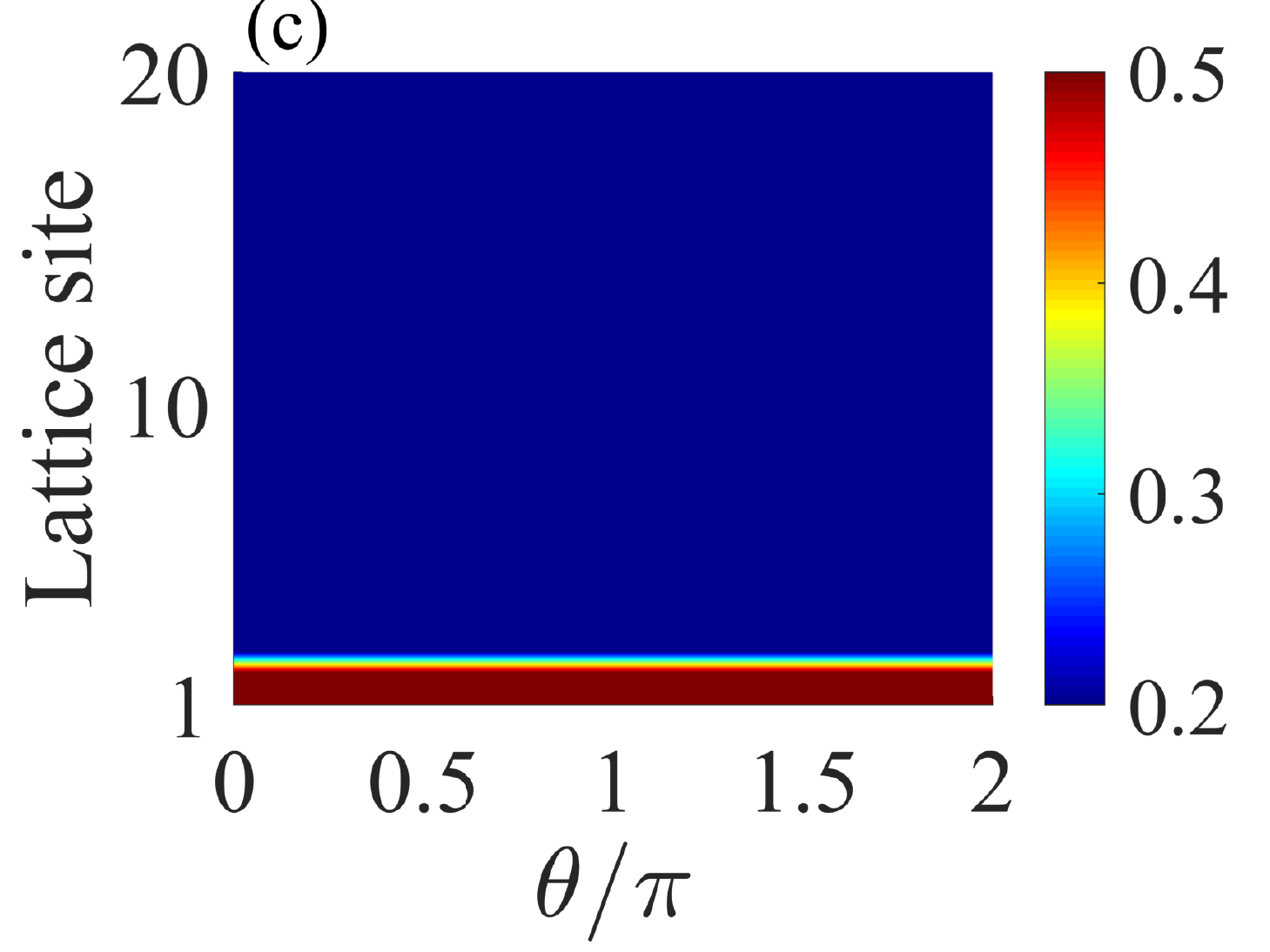}}
	\subfigure{\includegraphics[width=0.48\linewidth]{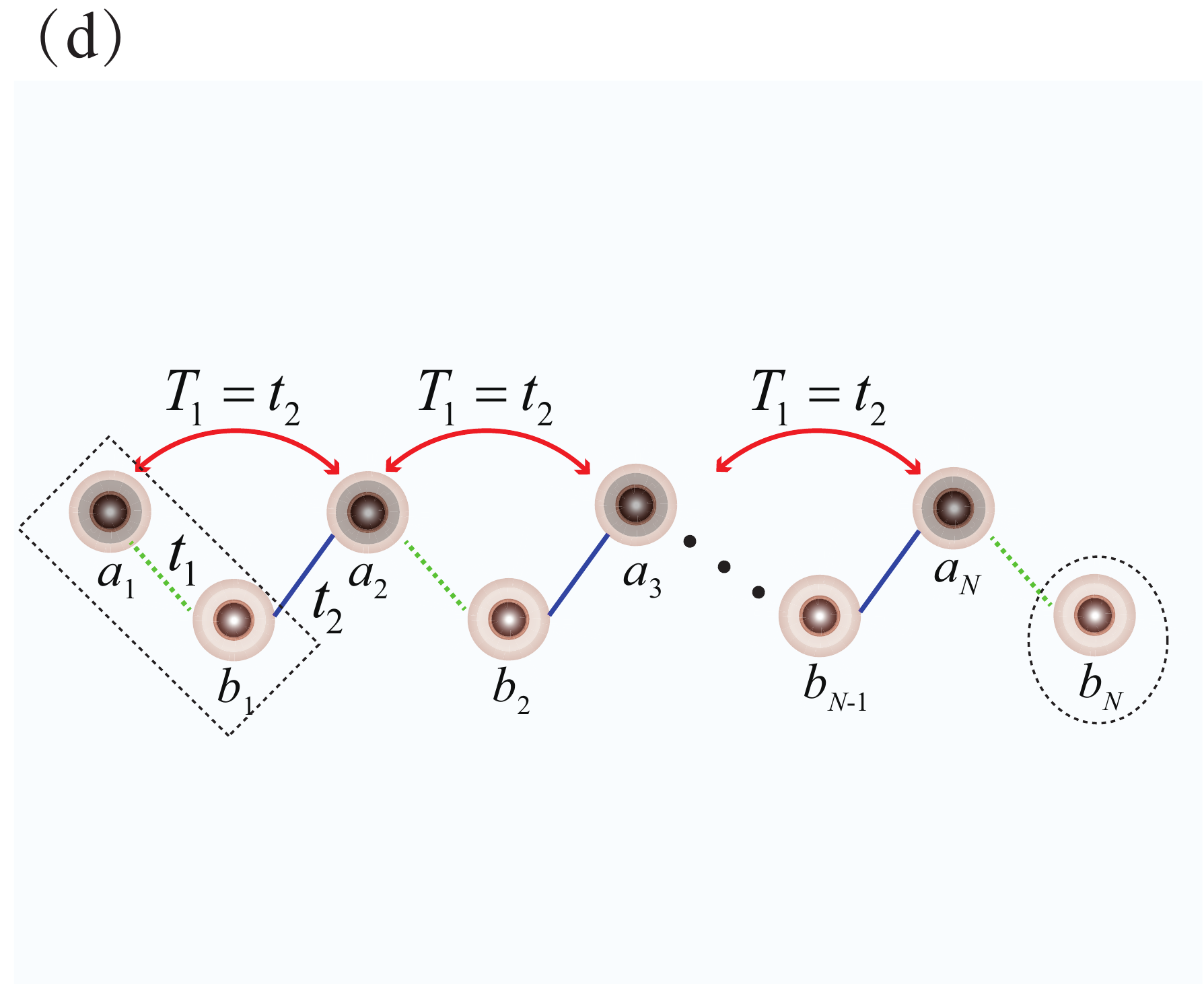}}
	\caption{The energy spectrum and distribution of states. (a) The energy spectrum of the SSH model with $V_{a}=V_{b}=T_{2}=0$, $t_{1}=1-\cos\theta$, $t_{2}=1+\cos\theta$, and $T_{1}=t_{2}=1+\cos\theta$. The energy gap has two separated gap states. The blue dashed line represents that the gap state is localized at right edge while the green line represents that the gap state is localized at first two sites with the same distributions. (b) The corresponding distribution of the upper gap state in (a). (c) The corresponding distribution of the lower gap state  in (a). (d) The schematic diagram of the two different kinds of distributions of gap states. The black dashed circle represents the right edge state, and the black dashed rectangular box represent the super-site induced by the equal $T_{1}=t_{2}$. Other parameter takes $N=10$.}\label{fig9}
\end{figure}
\begin{figure}
	\centering
	\includegraphics[width=1.0\linewidth]{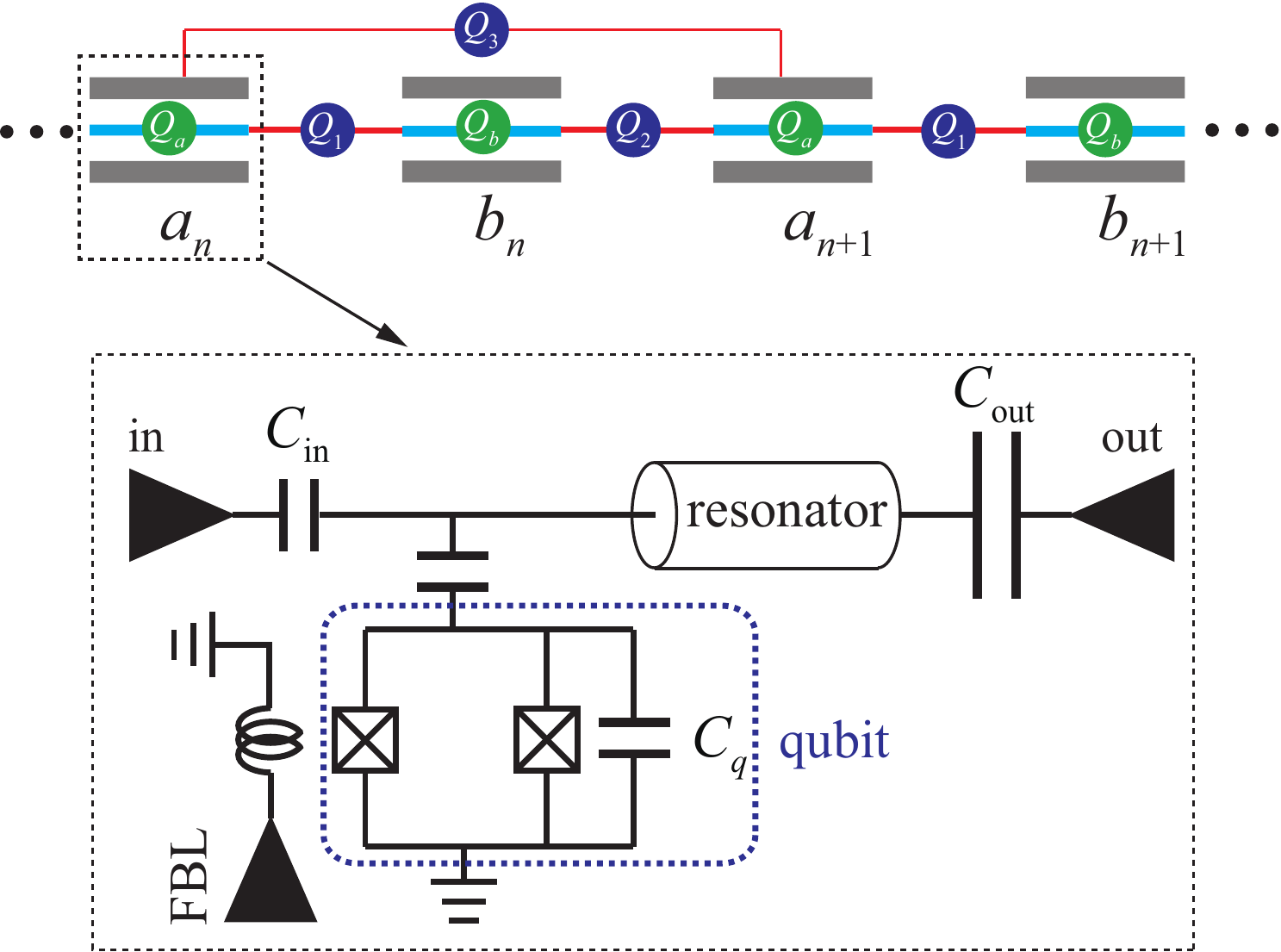}
	\caption{The realization of the topological beam splitter based on circuit-QED lattice. The up panel depicts the structure of the circuit-QED lattice with the NNN hoppings. Each resonator couples a superconducting qubit and the two resonators couple with each other via other superconducting qubits. The bottom panel shows the circuit structure of the coupling between the resonator and the qubit. The energy level space of the qubit can be tuned by the flux-bias line (FBL).}\label{fig10}
\end{figure}

\subsection{Theoretical analysis of the physical mechanisms for topological beam splitter}\label{sec.3B}
In the previous section, we have demonstrated that the topological beam splitter can be constructed via the topological gap state induced by the modulated on-site energy and NNN hoppings. Here we further reveal the internal physical mechanisms of the topological beam splitter. When the parameters in Eq.~(\ref{e06}) satisfy $V_{a}=V_{b}=T_{2}=0$, $t_{1}=1-\cos\theta$, $t_{2}=1+\cos\theta$, and $T_{1}=t_{2}=1+\cos\theta$, the modulated NNN hoppings added only on the odd sites make the energy spectrum deformed, as shown in Fig.~\ref{fig9}(a). The corresponding distributions of two gap states are shown in Figs.~\ref{fig9}(b) and~\ref{fig9}(c). The results indicate that the upper gap state in Fig.~\ref{fig9}(a) is always localized near the rightmost site when $\theta\in[0, 0.5\pi]\cup[1.5\pi, 2\pi]$, while the lower gap state in Fig.~\ref{fig9}(a) is always localized at the first two sites with the approximately same distribution when $\theta\in[0, 2\pi]$. The reason for the distribution of the upper gap state can be interpreted as following. When the parameter $\theta$ satisfies $\theta\in[0, 0.5\pi]\cup[1.5\pi, 2\pi]$, the NN hopping terms always satisfy $t_{1}<t_{2}$. The weaker $t_{1}$ makes the last site $b_{N}$ decoupled from the lattice chain, leading to the existence of the topological right edge state, as shown in Fig.~\ref{fig9}(d). 

On the contrary, the distribution of the lower green gap state is induced by the NNN hoppings added only on the odd sites. More specifically, when the NNN hoppings strength $T_{1}=t_{2}$ is added into the system, the identical NN hopping strength $t_{2}$ and the NNN hoppings strength $T_{1}$ make the first site $a_{1}$ and the second site $b_{1}$ have the same hopping probability towards the third site $a_{2}$, as shown in Fig.~\ref{fig9}(d). The same hopping probability towards the site $a_{2}$ leads the first two sites $a_{1}$ and $b_{1}$ to be equivalent to a super-site [black dashed rectangular box in Fig.~\ref{fig9}(d)]~\cite{Qi2020Topological}. At the same time, the rightmost site $b_{N}$ is not affected by the NNN hoppings strength $T_{1}=t_{2}$ since the NNN hoppings are only added on the odd sites. Therefore, the distribution of the upper gap state is always localized near the last site when $\theta\in[0, 0.5\pi]\cup[1.5\pi, 2\pi]$. The right edge state with the maximal distribution at the last site and the left edge state with the equal distribution at the first two sites provide the physical basis for constructing the topological beam splitter. 

Although the two gap states in Fig.~\ref{fig9}(a) provide the possibility for the implementation of topological beam splitter, both the two gap states have the same distribution at different regions of $\theta$, implying that the state transfer between different states cannot be achieved [the same as the case in Fig.~\ref{fig2}(b)]. Therefore, as demonstrated in section~\ref{sec.2A}, we need to introduce the staggered modulated on-site energy to transform the distributions of the two kinds of states. In this way, the joint effect between the staggered modulated on-site energy with $V_{a}=-V_{b}=-\sin\theta$ and the modulated NNN hoppings added only on the odd sites with $T_{1}=t_{2}=1+\cos\theta$ induces the topological channel for the realization of the state transfer between $|0,~0,~0,...,0,~1\rangle$ and $|0.5,~0.5,~0,...,0,~0\rangle$. Namely, the construction of topological beam splitter.   
\begin{figure}
	\centering
	\subfigure{\includegraphics[width=0.85\linewidth]{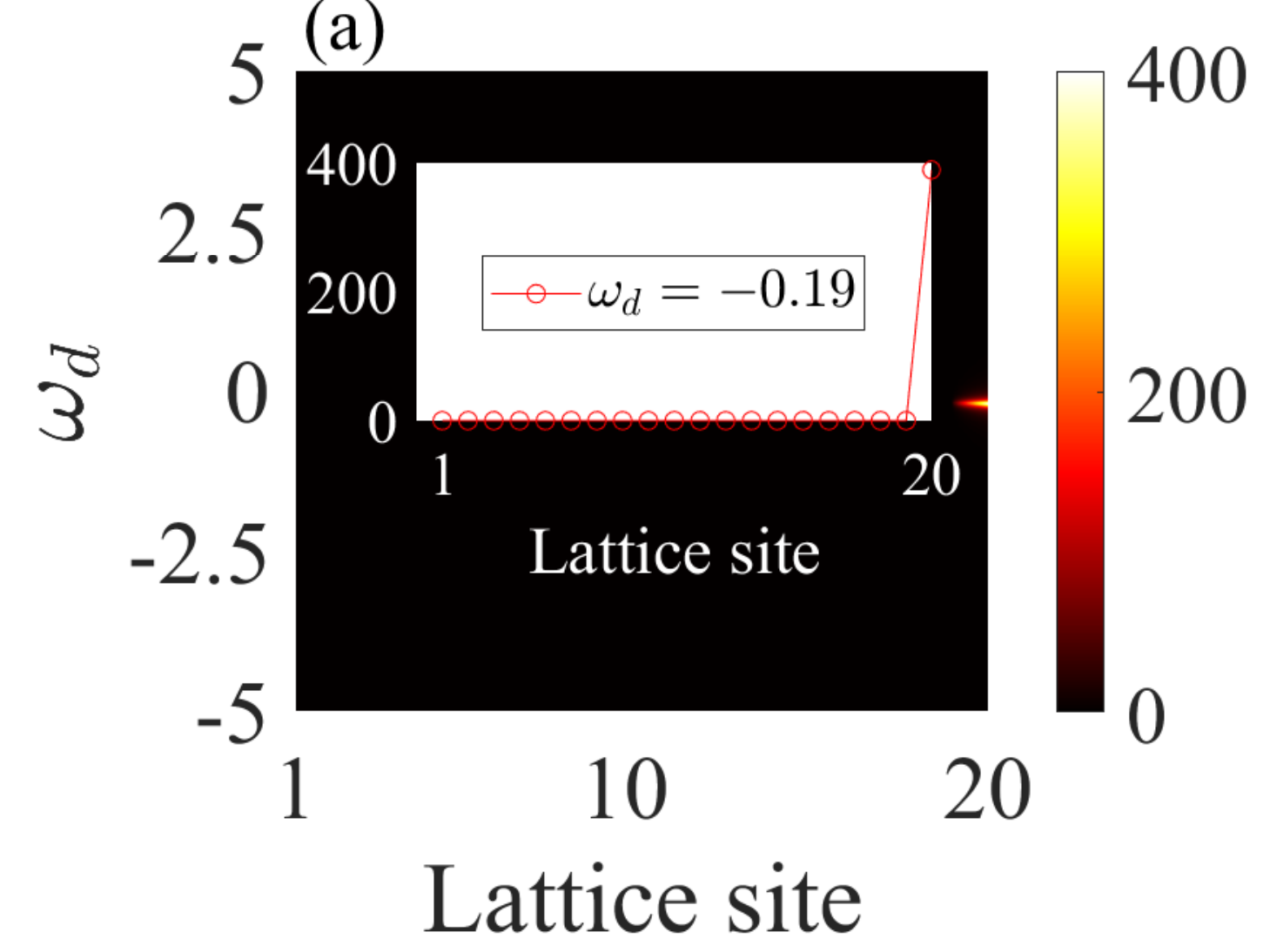}}
	
	\subfigure{\includegraphics[width=0.85\linewidth]{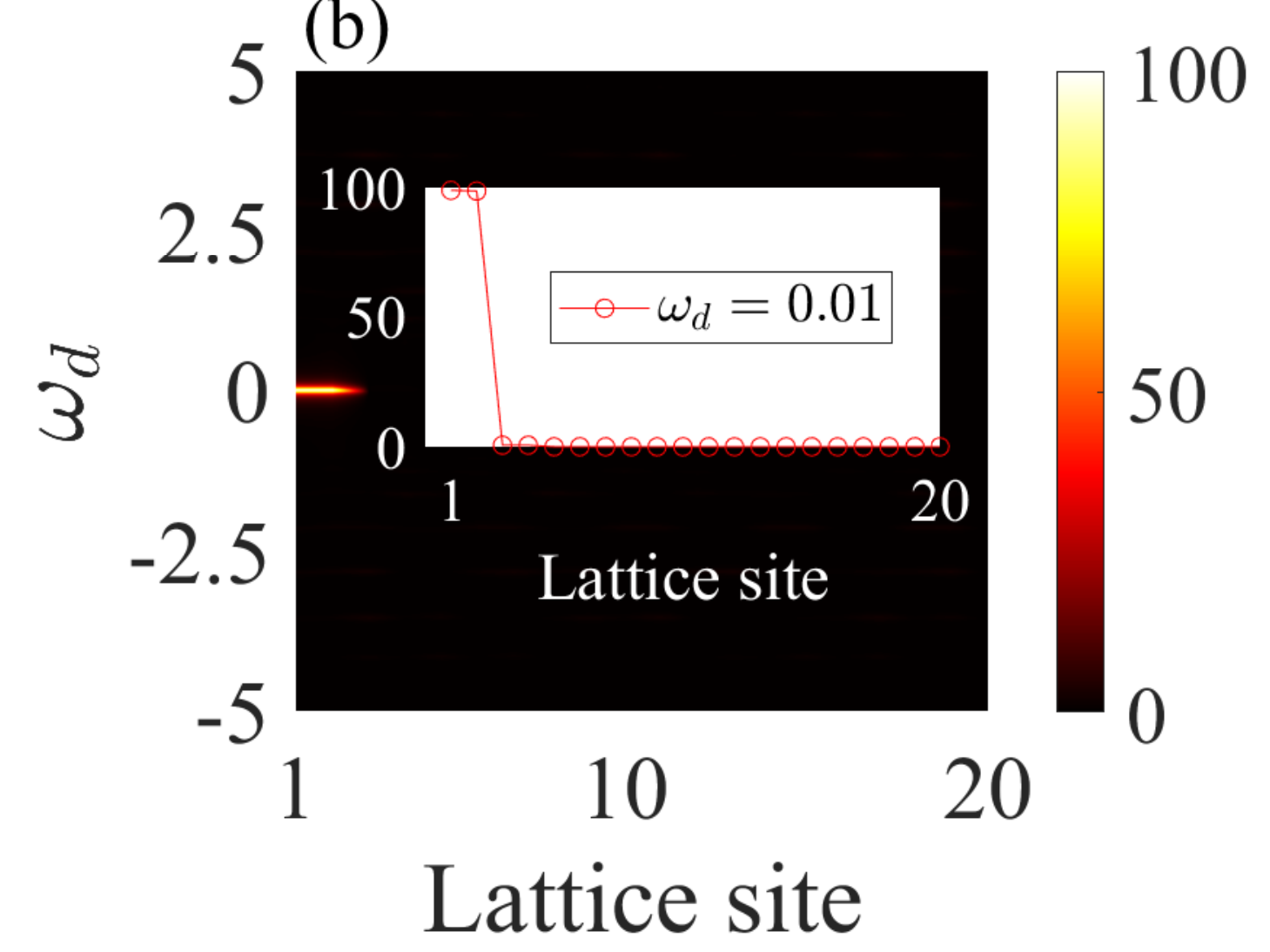}}
	\caption{The output detection spectrum. (a) The external driving excites the last resonator with $\Omega_{b_{N}}=1$. The photons mainly gather into the rightmost resonator with the resonant excitation. (b) The external driving excites the first resonator $\Omega_{a_{1}}=1$. The photons mainly gather into the first two resonators with the resonant excitation. Other parameters take $\theta=0.15$ and $N=10$.}\label{fig11}
\end{figure}

\subsection{Experimental realization of topological beam splitter}\label{sec.3C}
The beam splitter has many significant applications in the field of quantum optics. Benefiting from the rapid development in circuit-QED community, the circuit-QED system provides us an excellent experimental platform to realize the topological beam splitter proposed in our work. Here, we construct a circuit-QED lattice via arranging the transmission line resonator and the superconducting qubits in the space, as shown in Fig.~\ref{fig10}, in which each resonator $a_{n}$ ($b_{n}$) couples a superconducting qubits $Q_{a}$ ($Q_{b}$) with the coupling strength $g_{a}$ ($g_{b}$), the two adjacent resonators $a_{n}$ ($a_{n+1}$) and $b_{n}$ couple with each other via qubit $Q_{1}$ ($Q_{2}$) with the coupling strength $g_{1}$ ($g_{2}$), and the two resonators $a_{n}$ and $a_{n+1}$ couples with each other via qubit $Q_{3}$ with the coupling strength $g_{3}$. Under the dissipative regime~\cite{mei2018robust,mei2016Witnessing}, if all the qubits are initially prepared in their ground states, the effective Hamiltonian of the circuit-QED lattice system (after resetting the energy zero point with respect to the resonator detuning) is given by
\begin{eqnarray}\label{e07}
H&=&\sum_{n}-(\frac{g_{a}^{2}}{\Delta_{a}}+\frac{g_{1}^{2}}{\Delta_{1}}+\frac{g_{2}^{2}}{\Delta_{2}}+\frac{2g_{3}^{2}}{\Delta_{3}})a_{n}^{\dag}a_{n}\cr\cr
&&-(\frac{g_{b}^{2}}{\Delta_{b}}+\frac{g_{1}^{2}}{\Delta_{1}}+\frac{g_{2}^{2}}{\Delta_{2}})b_{n}^{\dag}b_{n}\cr\cr
&&-\left[\frac{g_{1}^{2}}{\Delta_{1}}a_{n}^{\dag}b_{n}+\frac{g_{2}^{2}}{\Delta_{2}}a_{n+1}^{\dag}b_{n}+\frac{g_{3}^{2}}{\Delta_{3}}a_{n+1}^{\dag}a_{n}+\mathrm{H.c.}\right],\cr&&
\end{eqnarray}   
where $\Delta_{i}=\omega_{i}-\omega_{d}$ ($i=a,b,1,2,3$) represents the qubit detuning between qubit level $\omega_{i}$ and driving frequency $\omega_{d}$. Note that the energy levels of the superconducting qubits in Fig.~\ref{fig10} are tunable via magnetic flux provided by a flux-bias line (FBL)~\cite{Schmidt2013Circuit}, making that the detuning of the qubits are tunable, which provides the basis for the periodically modulated terms. Generally, the typical qubit level can be modulated in the range of $100\mathrm{MHz}$ to $15\mathrm{GHz}$~\cite{Manucharyan2009Fluxonium,Manucharyan2012Evidence}, providing a considerable adjustability in experiment. For simplicity, we take the coupling strength $g_{i}$ ($i=a,b,1,2,3$) between the resonator and the qubit as $-g_{i}=1$ as the energy unit. Then, the periodically modulated on-site energy, NN hopping terms, and NNN hoppings can be easily realized by modulating the qubit detuning $\Delta_{i}$ originating from the qubit level via FBL. To construct the proposed topological beam splitter, a typical choice of the qubit detuning is that $\frac{1}{\Delta_{1}}=1-\cos\theta$, $\frac{1}{\Delta_{2}}=\frac{1}{\Delta_{3}}=1+\cos\theta$, $\frac{1}{\Delta_{a}}=-(4+2\cos\theta+\sin\theta)$, and $\frac{1}{\Delta_{b}}=-2+\sin\theta$. Under this set of parameter regime, the Hamiltonian in Eq.~(\ref{e07}) becomes
\begin{eqnarray}\label{e08}
H&=&\sum_{n}-\sin\theta a_{n}^{\dag}a_{n}+\sin\theta b_{n}^{\dag}b_{n}+\left[(1-\cos\theta)a_{n}^{\dag}b_{n}\right.\cr\cr
&&\left.+(1+\cos\theta)a_{n+1}^{\dag}b_{n}+(1+\cos\theta)a_{n+1}^{\dag}a_{n}\right]+\mathrm{H.c.}.\cr&&
\end{eqnarray}   
Obviously, the above Hamiltonian is the basis for realizing the topological beam splitter, and we can construct the topological beam splitter by controlling the external field adiabatically.

Another advantage of the construction of topological beam splitter based on circuit-QED lattice is that, the output photons can be detected by the resonator-based input-output formalism~\cite{mei2018robust,mei2016Witnessing}. If we use the external driving $H_{d}=\sum_{n}[\Omega_{a,n}a_{n}e^{i\omega_{d} t}+\Omega_{b,n}b_{n}e^{i\omega_{d} t}+\mathrm{H.c.}]$ ( $\Omega_{a,n}$ ($\Omega_{b,n}$) is the driving amplitude and $\omega_{d}$ is the driving frequency) to excite the circuit-QED lattice, we find that, under the steady state assumption, the input and the output photons can be detected. For example, when $\theta=0.15\pi$, we use the external driving to excite the rightmost resonator with a certain range of driving frequency. The corresponding output detection spectrum is shown in Fig.~\ref{fig11}(a), in which the numerical results show that, when the driving frequency reaches resonant with the red gap state in Fig.~\ref{fig6}, the photons mainly gather into the last resonator, which is consistent with the input state. Similarly, when we use the external driving to excite the first resonator or the second resonator with a certain range of driving frequency, we find that the photons mainly gather into the first two resonators with the same magnitude, as shown in Fig.~\ref{fig11}(b). The results reveal that we also can realize the detection of the output photons. Therefore, our scheme provides the possibility towards the realization of the topological beam splitter based on circuit-QED lattice system, which greatly expands the potential applications of topological matter in quantum information processing.

\section{Conclusions}\label{sec.4}
In conclusion, we have shown that the state transfer between the left and right edge states in the even-size SSH chain can be realized via the topological channels induced by the staggered periodic on-site potentials or the staggered periodic NNN hoppings. We find that the staggered periodic on-site potentials or the NNN hoppings lead to the complete splitting of the initial degenerate zero energy modes, which further facilitates the separation of the topological left and right edge states. In this way, the state transfers between the first $a$-type and the last $b$-type site can be achieved under the adiabatic evolution condition. Moreover, we introduce the staggered periodic on-site potentials and the periodic NNN hoppings added only on the odd sites simultaneously to induce a special edge channel, in which the right edge state can be transferred to the first two sites with the equal distribution. This property makes that the present topological channel has many potential applications in topological beam splitter device. Furthermore, we  demonstrate that, based on circuit-QED lattice, the topological beam splitter can be achieved under the current experimental conditions. Our work opens up a new way for the realization of topological quantum information processing and provides a new path towards the engineering of new type of quantum optical device.

\begin{center}
{\bf{ACKNOWLEDGMENTS}}
\end{center}
This work was supported by the National Natural Science Foundation of China under Grant Nos.
61822114, 11874132, 61575055, and 11575048, and the Project of Jilin Science and Technology Development for Leading Talent of Science and Technology Innovation in Middle and Young and Team Project under Grant No. 20160519022JH.

%\bibliographystyle{apsjnl}
%\bibliography{Reference}

\begin{thebibliography}{10}
	\newcommand{\enquote}[1]{``#1''}
	
	\bibitem{hasan2010colloquium}
	M.~Z. Hasan and C.~L. Kane, Colloquium: topological insulators, Rev. Mod. Phys.
	\textbf{82}, 3045 (2010).
	
	\bibitem{qi2011topological}
	X.~L. Qi and S.~C. Zhang, Topological insulators and superconductors, Rev. Mod.
	Phys. \textbf{83}, 1057 (2011).
	
	\bibitem{chiu2016classification}
	C.~K. Chiu, J.~C. Teo, A.~P. Schnyder, and S.~Ryu, Classification of
	topological quantum matter with symmetries, Rev. Mod. Phys. \textbf{88},
	035005 (2016).
	
	\bibitem{bansil2016colloquium}
	A.~Bansil, H.~Lin, and T.~Das, Colloquium: Topological band theory, Rev. Mod.
	Phys. \textbf{88}, 021004 (2016).
	
	\bibitem{matsuura2010momentum}
	S.~Matsuura and S.~Ryu, Momentum space metric, nonlocal operator, and
	topological insulators, Phys. Rev. B \textbf{82}, 245113 (2010).
	
	\bibitem{wray2011topological}
	L.~A. Wray, S.~Y. Xu, Y.~Xia, D.~Hsieh, A.~V. Fedorov, Y.~San~Hor, R.~J. Cava,
	A.~Bansil, H.~Lin, and M.~Z. Hasan, A topological insulator surface under
	strong coulomb, magnetic and disorder perturbations, Nat. Phys. \textbf{7},
	32 (2011).
	
	\bibitem{xu2006stability}
	C.~Xu and J.~E. Moore, Stability of the quantum spin hall effect: Effects of
	interactions, disorder, and $Z_{2}$ topology, Phys. Rev. B \textbf{73}, 045322
	(2006).
	
	\bibitem{malki2017tunable}
	M.~Malki and G.~Uhrig, Tunable edge states and their robustness towards
	disorder, Phys. Rev. B \textbf{95}, 235118 (2017).
	
	\bibitem{xiao2017observation}
	L.~Xiao, X.~Zhan, Z.~Bian, K.~Wang, X.~Zhang, X.~Wang, J.~Li, K.~Mochizuki,
	D.~Kim, N.~Kawakami \emph{et~al.}, Observation of topological edge states in
	parity--time-symmetric quantum walks, Nat. Phys. \textbf{13}, 1117--1123
	(2017).
	
	\bibitem{brouwer2011topological}
	P.~W. Brouwer, M.~Duckheim, A.~Romito, and F.~von Oppen, Topological
	superconducting phases in disordered quantum wires with strong spin-orbit
	coupling, Phys. Rev. B \textbf{84}, 144526 (2011).
	
	\bibitem{alicea2011non}
	J.~Alicea, Y.~Oreg, G.~Refael, F.~Von~Oppen, and M.~P. Fisher, Non-abelian
	statistics and topological quantum information processing in 1D wire
	networks, Nat. Phys. \textbf{7}, 412 (2011).
	
	\bibitem{duclos2010fast}
	G.~Duclos-Cianci and D.~Poulin, Fast decoders for topological quantum codes,
	Phys. Rev. Lett. \textbf{104}, 050504 (2010).
	
	\bibitem{sarma2015majorana}
	S.~D. Sarma, M.~Freedman, and C.~Nayak, Majorana zero modes and topological
	quantum computation, npj Quantum Inf. \textbf{1}, 15001 (2015).
	
	\bibitem{stern2013topological}
	A.~Stern and N.~H. Lindner, Topological quantum computation from basic
	concepts to first experiments, Science \textbf{339}, 1179--1184 (2013).
	
	\bibitem{dlaska2017robust}
	C.~Dlaska, B.~Vermersch, and P.~Zoller, Robust quantum state transfer via
	topologically protected edge channels in dipolar arrays, Quant. Sci. Technol. \textbf{2}, 015001 (2017).
	
	\bibitem{aasen2016milestones}
	D.~Aasen, M.~Hell, R.~V. Mishmash, A.~Higginbotham, J.~Danon, M.~Leijnse, T.~S.
	Jespersen, J.~A. Folk, C.~M. Marcus, K.~Flensberg \emph{et~al.}, Milestones
	toward majorana-based quantum computing, Phys. Rev. X \textbf{6}, 031016
	(2016).
	
	\bibitem{tewari2007quantum}
	S.~Tewari, S.~D. Sarma, C.~Nayak, C.~Zhang, and P.~Zoller, Quantum computation
	using vortices and majorana zero modes of a $p_{x}+ip_{y}$ superfluid of
	fermionic cold atoms, Phys. Rev. Lett. \textbf{98}, 010506 (2007).
	
	\bibitem{su1979solitons}
	W.~Su, J.~Schrieffer, and A.~J. Heeger, Solitons in polyacetylene, Phys. Rev.
	Lett. \textbf{42}, 1698 (1979).
	
	\bibitem{takayama1980continuum}
	H.~Takayama, Y.~R. Lin-Liu, and K.~Maki, Continuum model for solitons in
	polyacetylene, Phys. Rev. B \textbf{21}, 2388 (1980).
	
	\bibitem{fradkin1983phase}
	E.~Fradkin and J.~E. Hirsch, Phase diagram of one-dimensional electron-phonon
	systems. i. the su-schrieffer-heeger model, Phys. Rev. B \textbf{27}, 1680
	(1983).
	
	\bibitem{jackiw1976solitons}
	R.~Jackiw and C.~Rebbi, Solitons with fermion number $1/2$, Phys. Rev. D
	\textbf{13}, 3398 (1976).
	
	\bibitem{heeger1988solitons}
	A.~J. Heeger, S.~Kivelson, J.~Schrieffer, and W.~P. Su, Solitons in conducting
	polymers, Rev. Mod. Phys. \textbf{60}, 781 (1988).
	
	\bibitem{meier2016observation}
	E.~J. Meier, F.~A. An, and B.~Gadway, Observation of the topological soliton
	state in the Su-Schrieffer-Heeger model, Nat. Commun. \textbf{7}, 13986
	(2016).
	
	\bibitem{ganeshan2013topological}
	S.~Ganeshan, K.~Sun, and S.~D. Sarma, Topological zero-energy modes in gapless
	commensurate aubry-andr{\'e}-harper models, Phys. Rev. Lett. \textbf{110},
	180403 (2013).
	
	\bibitem{li2014topological}
	L.~Li, Z.~Xu, and S.~Chen, Topological phases of generalized
	Su-Schrieffer-Heeger models, Phys. Rev. B \textbf{89}, 085111 (2014).
	
	\bibitem{zhang2018topological}
	D.~W. Zhang, Y. Q. Zhu, Y.~Zhao, H.~Yan, and S.~L. Zhu, Topological quantum
	matter with cold atoms, Adv. Phys. \textbf{67}, 253--402 (2018).
	
	\bibitem{cheng2015topologically}
	Q.~Cheng, Y.~Pan, Q.~Wang, T.~Li, and S.~Zhu, Topologically protected interface
	mode in plasmonic waveguide arrays, Laser Photon. Rev. \textbf{9},
	392--398 (2015).
	
	\bibitem{longhi2013zak}
	S.~Longhi, Zak phase of photons in optical waveguide lattices, Opt. Lett.
	\textbf{38}, 3716--3719 (2013).
	
	\bibitem{ke2019topological}
	S.~Ke, D.~Zhao, J.~Liu, Q.~Liu, Q.~Liao, B.~Wang, and P.~Lu, Topological bound
	modes in anti-PT-symmetric optical waveguide arrays, Opt. Express
	\textbf{27}, 13858--13870 (2019).
	
	\bibitem{groning2018engineering}
	O.~Gr{\"o}ning, S.~Wang, X.~Yao, C.~A. Pignedoli, G.~B. Barin, C.~Daniels,
	A.~Cupo, V.~Meunier, X.~Feng, A.~Narita \emph{et~al.}, Engineering of robust
	topological quantum phases in graphene nanoribbons, Nature \textbf{560}, 209
	(2018).
	
	\bibitem{ribeiro2015transport}
	L.~A. Ribeiro~Jr, W.~F. da~Cunha, A.~L. d.~A. Fonseca, G.~M. e~Silva, and
	S.~Stafstr\"{o}m, Transport of polarons in graphene nanoribbons, J. Phys. Chem. Lett. \textbf{6}, 510--514 (2015).
	
	\bibitem{da2015impurity}
	W.~F. da~Cunha, L.~A.~R. Junior, A.~L. de~Almeida~Fonseca, R.~Gargano, and
	G.~M. e~Silva, Impurity effects on polaron dynamics in graphene nanoribbons,
	Carbon \textbf{91}, 171--177 (2015).
	
	\bibitem{cao2017topological}
	T.~Cao, F.~Zhao, and S.~G. Louie, Topological phases in graphene nanoribbons:
	junction states, spin centers, and quantum spin chains, Phys. Rev. Lett.
	\textbf{119}, 076401 (2017).
	
	\bibitem{engelhardt2017topologically}
	G.~Engelhardt, M.~Benito, G.~Platero, and T.~Brandes, Topologically enforced
	bifurcations in superconducting circuits, Phys. Rev. Lett. \textbf{118},
	197702 (2017).
	
	\bibitem{paraoanu2014recent}
	G.~Paraoanu, Recent progress in quantum simulation using superconducting
	circuits, J. Low Temp. Phys. \textbf{175}, 633--654 (2014).
	
	\bibitem{qi2018simulation}
	L.~Qi, Y.~Xing, J.~Cao, X.~X. Jiang, C.~S. An, A.~D. Zhu, S.~Zhang, and H.~F.
	Wang, Simulation and detection of the topological properties of a modulated
	Rice-Mele model in a one-dimensional circuit-QED lattice, Sci. China-Phys. Mech. Astron. \textbf{61}, 080313 (2018).
	
	\bibitem{esmann2018topological}
	M.~Esmann and N.~Lanzillotti-Kimura, A topological view on optical and phononic
	fabry--perot microcavities through the Su-Schrieffer-Heeger model, Appl. Sci. \textbf{8}, 527 (2018).
	
	\bibitem{xing2018controllable}
	Y.~Xing, L.~Qi, J.~Cao, D.~Y. Wang, C.~H. Bai, W.~X. Cui, H.~F. Wang, A.~D.
	Zhu, and S.~Zhang, Controllable photonic and phononic edge localization via
	optomechanically induced kitaev phase, Opt. Express \textbf{26}, 16250--16264
	(2018).
	
	\bibitem{xu2018generalized}
	X.~W. Xu, Y.~J. Zhao, H.~Wang, A.~X. Chen, and Y.~X. Liu, Generalized
	Su-Schrieffer-Heeger model in one dimensional optomechanical arrays, arXiv:1807.07880  (2018).
	
	\bibitem{qi2020controllable}
	L. Qi, S. Liu, S. Zhang, and H. F. Wang, Controllable double quantum state transfers by one topological channel in a frequency-modulated optomechanical array, Opt. Letters \textbf{45}, 2018-2021 (2020).
	
	\bibitem{grusdt2013topological}
	F.~Grusdt, M.~H{\"o}ning, and M.~Fleischhauer, Topological edge states in the
	one-dimensional superlattice bose-hubbard model, Phys. Rev. Lett.
	\textbf{110}, 260405 (2013).
	
	\bibitem{yao2018edge}
	S.~Yao and Z.~Wang, Edge states and topological invariants of non-hermitian
	systems, Phys. Rev. Lett. \textbf{121}, 086803 (2018).
	
	\bibitem{di2016two}
	M.~Di~Liberto, A.~Recati, I.~Carusotto, and C.~Menotti, Two-body physics in the
	Su-Schrieffer-Heeger model, Phys. Rev. A \textbf{94}, 062704 (2016).
	
	\bibitem{kitagawa2010exploring}
	T.~Kitagawa, M.~S. Rudner, E.~Berg, and E.~Demler, Exploring topological phases
	with quantum walks, Phys. Rev. A \textbf{82}, 033429 (2010).
	
	\bibitem{asboth2013bulk}
	J.~K. Asb{\'o}th and H.~Obuse, Bulk-boundary correspondence for chiral
	symmetric quantum walks, Phys. Rev. B \textbf{88}, 121406 (2013).
	
	\bibitem{ezawa2019electric}
	M.~Ezawa, Electric-circuit simulation of the schr{\"o}dinger equation and
	non-Hermitian quantum walks, Phys. Rev. B \textbf{100}, 165419 (2019).
	
	\bibitem{lieu2018topological}
	S.~Lieu, Topological phases in the non-Hermitian Su-Schrieffer-Heeger model,
	Phys. Rev. B \textbf{97}, 045106 (2018).
	
	\bibitem{zhu2014pt}
	B.~Zhu, R.~L{\"u}, and S.~Chen, $\mathcal{P}$$\mathcal{T}$-symmetry in the non-Hermitian
	Su-Schrieffer-Heeger model with complex boundary potentials, Phys. Rev. A
	\textbf{89}, 062102 (2014).
	
	\bibitem{kunst2018biorthogonal}
	F.~K. Kunst, E.~Edvardsson, J.~C. Budich, and E.~J. Bergholtz, Biorthogonal
	bulk-boundary correspondence in non-Hermitian systems, Phys. Rev. Lett.
	\textbf{121}, 026808 (2018).
	
	\bibitem{wang2013topological}
	L.~Wang, M.~Troyer, and X.~Dai, Topological charge pumping in a one-dimensional
	optical lattice, Phys. Rev. Lett. \textbf{111}, 026802 (2013).
	
	\bibitem{lohse2016thouless}
	M.~Lohse, C.~Schweizer, O.~Zilberberg, M.~Aidelsburger, and I.~Bloch, A
	thouless quantum pump with ultracold bosonic atoms in an optical
	superlattice, Nat. Phys. \textbf{12}, 350 (2016).
	
	\bibitem{mei2018topology}
	F.~Mei, G.~Chen, L.~Tian, S.~L. Zhu, and S.~Jia, Topology-dependent quantum
	dynamics and entanglement-dependent topological pumping in superconducting
	qubit chains, Phys. Rev. A \textbf{98}, 032323 (2018).
	
	\bibitem{hu2019dispersion}
	S.~Hu, Y.~Ke, Y.~Deng, and C.~Lee, Dispersion-suppressed topological thouless pumping, Phys. Rev. B \textbf{100}, 064302 (2019).
	
	\bibitem{xie2019topological}
	D.~Xie, W.~Gou, T.~Xiao, B.~Gadway, and B.~Yan, Topological characterizations
	of an extended Su-Schrieffer-Heeger model, npj Quantum Inf. \textbf{5}, 1
	(2019).
	
	\bibitem{rakovszky2017detecting}
	T.~Rakovszky, J.~K. Asb{\'o}th, and A.~Alberti, Detecting topological
	invariants in chiral symmetric insulators via losses, Phys. Rev. B
	\textbf{95}, 201407 (2017).
	
	\bibitem{velasco2017realizing}
	C.~G. Velasco and B.~Paredes, Realizing and detecting a topological insulator
	in the A$\mathrm{\Rmnum{3}}$ symmetry class, Phys. Rev. Lett. \textbf{119}, 115301 (2017).
	
	\bibitem{mei2018robust}
	F.~Mei, G.~Chen, L.~Tian, S.~L. Zhu, and S.~Jia, Robust quantum state transfer
	via topological edge states in superconducting qubit chains, Phys. Rev. A
	\textbf{98}, 012331 (2018).
	
	\bibitem{asboth2016ashort}
	J. K. Asb\'{o}th, L. Oroszl\'{a}ny, and A. P\'{a}lyi, A short course on topological insulators, Lecture notes in physics, \textbf{919}, 87 (2016).
	
	\bibitem{Jonckheere2017Hanbury}
    T. Jonckheere, J. Rech, A. Zazunov, R. Egger, and T. Martin, Hanbury Brown and Twiss noise correlations in a topological superconductor beam splitter, Phys. Rev. B, \textbf{95}, 054514 (2017)	
	
	\bibitem{Wang2017Topologically}
    X. S. Wang, Y. Su, and X. R. Wang, Topologically protected unidirectional edge spin waves and beam splitter, Phys. Rev. B, \textbf{95}, 014435 (2017)		
	
	\bibitem{Hammer2013Dynamics}
    R. Hammer and W. P\"{o}tz, Dynamics of domain-wall Dirac fermions on a topological insulator: A chiral fermion beam splitter, Phys. Rev. B, \textbf{88}, 235119 (2013)		
	
    \bibitem{Macneille1946}
	S. M. Macneille, U.S. Patent No.2, 403, 731. Washington, DC: U.S. Patent and Trademark Office (1946)
		
	\bibitem{Schonbrun2006Yamashita}
	E. Schonbrun, Q. Wu, W. Park, T. Yamashita, and C. J. Summers, Polarization beam splitter based on a photonic crystal heterostructure, Opt. Lett. \textbf{31}, 3104-3106 (2006)
	
	\bibitem{Chen2004Photonic}
	C. C. Chen, H. D. Chien, and P. G. Luan, Photonic crystal beam splitters, Appl. Opt. \textbf{43}, 6187-6190 (2004)
	
	\bibitem{Donatella2000Beam}
    C. Donatella, H. Bj\"{o}rn, F. Ron, M. Thomas, and S. J\"{o}rg, Beam Splitter for Guided Atoms, Phys. Rev. Lett. \textbf{85}, 5483 (2000)
    
    \bibitem{Qi2020Topological}
    L. Qi, Y. Xing, S. Liu, S. Zhang, and H. F. Wang, Topological phase induced by distinguishing parameter regimes in a cavity optomechanical system with multiple mechanical resonators, Phys. Rev. A \textbf{101}, 052325 (2020)
    
	\bibitem{mei2016Witnessing}
	F. Mei, Z. Y. Xue, D. W. Zhang, L. Tian,  C. Lee, and S. L. Zhu, Witnessing topological Weyl semimetal phase in a minimal circuit-QED lattice, Quantum Sci. Technol. \textbf{1}, 015006 (2016)	
	
	\bibitem{Schmidt2013Circuit}
	S. Schmidt, and J. Koch,  Circuit QED lattices: towards quantum simulation with superconducting circuits, Ann. Phys. (Berlin) \textbf{525}, 395-412 (2013)
	
	\bibitem{Manucharyan2009Fluxonium}
	V. E. Manucharyan, J. Koch, L. I. Glazman, and M. H. Devoret, Fluxonium: Single cooper-pair circuit free of charge offsets, Science \textbf{326}, 113-116 (2009)
	
	\bibitem{Manucharyan2012Evidence}
	V. E. Manucharyan, N. A. Masluk, A. Kamal, J. Koch, L. I. Glazman, and M. H. Devoret, Evidence for coherent quantum phase slips across a Josephson junction array,  Phys. Rev. B \textbf{85}, 024521 (2012).
	
\end{thebibliography}

\end{document}